# T Duality of Perturbative Characters for Closed Bosonic and Type II Strings Theories.

**Eugène CREMMER**

**Jean-Loup GERVAIS**

Laboratoire de Physique Théorique de l'École Normale Supérieure[1],
24 rue Lhomond, 75231 Paris CEDEX 05, France.

## Abstract

A systematic study of duality properties of string characters associated with the transverse space-time rotations is undertaken. Although yet restricted to the perturbative spectrum, our work shows that these characters, already considered earlier as a book-keeping device for the free string quantum numbers, are useful tools in connection with the current developments of string duality ideas. In particular, the $O(8)$ triality properties of the characters for critical superstrings, show that there exists a third formulation (unnoticed so far to our knowledge) equivalent to, but different from the ones of Neveu-Schwarz-Ramond, and Green-Schwarz. For the bosonic modes, modular invariance requires that one also takes the trace over the total momentum, in contrast to what was done earlier. Although we do not consider supercharacters, the consequences of supersymmetry are neatly derived at once for all massive states by factorising the character of the long SUSY multiplet.

[1]Unité Propre du Centre National de la Recherche Scientifique, associée à l'École Normale Supérieure et à l'Université de Paris-Sud.

# Contents

# 1 Introduction

In string theories, the torus partition functions, which are of the type

$$P(q_0) = \mathrm{Tr}\left\{q_0^{L_0-1}\bar{q}_0^{\tilde{L}_0-1}\right\}$$

are familiar objects which count the degeneracy of each mass level. Their modular properties are well known. In light-cone approach, the trace is taken over the $D-2$ transverse string modes which obviously span a reducible representation of the transverse rotations $O(D-2)$. The point of the present work is to study the associated characters, which are natural generalisations of the partition functions just mentionned. We shall thus follow a path already travelled earlier[1], [2]. But we have undertaken to study the duality properties of these characters which is a novel viewpoint to our knowledge.

In general, let $G$ be a Lie group, with elements noted $g$. As is well known, it is useful to consider the trace

$$\chi_{\text{rep.}} = \mathrm{Tr}_{\text{rep.}}\left\{g\right\}$$

where $g$ is a particular element in a given representation. For the case we will consider, we may assume that $g$ is diagonalisable, and write

$$g = u\left(e^{2\pi i\sum_{\mu=1}^{N} v_\mu \mathcal{H}_\mu}\right)u^{-1},$$

where $N$ is the rank of $G$, and $\mathcal{H}_\mu$ are generators of the commutative Cartan subalgebra of $G$. Therefore

$$\chi_{\text{rep.}} = \chi_{\text{rep.}}(v_1,\ldots,v_N) = \mathrm{Tr}_{\text{rep.}}\left\{e^{2\pi i\sum_{\mu=1}^{N} v_\mu \mathcal{H}_\mu}\right\}.$$

Characters are functions of $N$ variables that completely characterise the representation. They are very nice tools, since being given by a trace, they do not depend upon the explicit realisation. It is hardly necessary to emphasize their importance in many fields of mathematics and theoretical physics. In string theories, they have so far been used as a powerful tool to study the representation content of string theories[2]. We will see that to study their duality properties gives a novel insight. In particular the $O(8)$ triality of superstring characters, will lead us, as a by-product, to a "third" formalism equivalent to but different from the NS-R and GS ones.

Although the present work will remain strictly at the perturbative level for standard string theories, our motivation is the hope to go far beyond. Our future aim is to use characters as a book-keeping device of latest extensions/generalisations of string theories. Indeed, they provide more precise informations about the theories than partition functions, but are still insensitive to the details of the dynamics so that we may hope to determine them exactly even non perturbatively. Being independent from the explicit realisations of multiplets, they allow in principle to handle perturbative and non perturbative states on the same footing. Another motivation is that the modular transformations we will study are notoriously rigid properties which mix all the states. and thus could give a handle on non BPS states. In this connection, let us note that, we define modular invariant characters, in contrast with the previous studies [2]. This is achieved by suitably integrating over the transverse total momentum, as explained in section 3.

# 2 Chiral bosonic characters with zero transverse momenta.

## 2.1 Calculation

A Lorentz vector $V$ in $D$ dimensions is represented by

$$V^\mu,\ \mu = 1, \dots D-2, \quad V^\pm = (V^0 \pm V^{D-1})/2.$$

In our perturbative situation we sum up[2] over physical free string states with $p^\mu = 0$. We use the light-cone formalism and only introduce string modes with transverse components. There is an obvious invariance by the

[2]repeating the discussion of refs.[2] for completeness,

corresponding $O(D-2)$ group. We shall only consider critical strings so that $D$ is even, and we let $N=(D-2)/2$. The string states span a reducible representation of $O(D-2)\equiv O(2N)=D_N$. The oscillator part[3] of the $D_N$ generators is[4]

$$\mathcal{M}^{\mu\nu}=\sum_{n\geq 1}\frac{1}{n}(\alpha^{\mu}_{-n}\alpha^{\nu}_{n}-\alpha^{\nu}_{-n}\alpha^{\mu}_{n}) \tag{2.1}$$

At this point we only consider one type of movers. In order to make contact with the usual Cartan basis of $D_N$ (see appendix A) it is useful to split the coordinates in two sets with indices $\pm\mu$, with $\mu=1,\ldots,N$. It is convenient to introduce, for $\mu>0$

$$\zeta^{\mu}_{r}=\frac{1}{\sqrt{2}}(\alpha^{\mu}_{r}-i\alpha^{-\mu}_{r}),\quad \zeta^{-\mu}_{r}=\frac{1}{\sqrt{2}}(\alpha^{\mu}_{r}+i\alpha^{-\mu}_{r}). \tag{2.2}$$

Clearly, for arbitary $\mu$, $\nu$,

$$\left[\zeta^{\mu}_{m},\zeta^{\nu}_{n}\right]=m\delta_{\mu,\,-\nu}\delta_{m,\,-n},\quad (\zeta^{\mu}_{m})^{+}=\zeta^{-\mu}_{-m}. \tag{2.3}$$

A set of commuting rotations is obtained by considering $\mathcal{M}^{\mu,-\mu}$ $\mu=1,\ldots,N$. The corresponding Cartan generators are to be defined as $\mathcal{H}_\mu=-i\mathcal{M}^{\mu,-\mu}$, that is

$$\mathcal{H}_{\mu}=\sum_{n\geq 1}\frac{1}{n}(\zeta^{-\mu}_{-n}\zeta^{\mu}_{n}-\zeta^{\mu}_{-n}\zeta^{-\mu}_{n}),\quad 1\leq\mu\leq N \tag{2.4}$$

These generators obviously commute also with the Virasoro generator of transverse modes

$$L_0=\sum_{n\geq 1}\left(\zeta^{-\mu}_{-n}\zeta^{\mu}_{n}+\zeta^{\mu}_{-n}\zeta^{-\mu}_{n}\right).$$

So one defines

$$\chi(q_0\,|\,q_1,\,\ldots,\,q_N)=\mathrm{Tr}\left\{q_0^{L_0}\prod_{\mu=1}^{N}q_\mu^{\mathcal{H}_\mu}\right\}$$

$$\equiv\prod_{n\geq 1}\mathrm{Tr}_n\left\{\prod_{\mu=1}^{N}q_0^{(\zeta^{-\mu}_{-n}\zeta^{\mu}_{n}+\zeta^{\mu}_{-n}\zeta^{-\mu}_{n})}\,q_\mu^{\frac{1}{n}(\zeta^{-\mu}_{-n}\zeta^{\mu}_{n}-\zeta^{\mu}_{-n}\zeta^{-\mu}_{n})}\right\}. \tag{2.5}$$

[3]Here we are not concerned about the zero modes which are not summed over. This will be done in section 3.

[4]Our conventions coincide with the ones of Green, Schwarz, Witten[3].

A straightforward computation gives

$$\chi(q_0 \,|\, q_1,\, \ldots,\, q_N) = \prod_{\mu=1}^{N} \prod_{n\geq 1} \frac{1}{1-q_0^n q_\mu} \frac{1}{1-(q_0^n/q_\mu)}. \tag{2.6}$$

## 2.2 Connection with the universal character of the linear group:

The representation content of this string spectrum is exhibited by expanding this character as a sum of irreducible characters of $D_N$. Moreover one knows that the massive string states actually span representations of $O(N+1) \equiv B_N$, and it should be possible to recombine sums of irreducible $D_N$ characters into irreducible $B_N$ ones[5]. This has been discussed in refs.[2] using the orthogonality properties of irreducible characters. Without going through a full discussion we next describe other powerful mathematical tools[4], [5], [7], [8], which are handy in this connection, and should be useful in the future. First, there is a direct connection between Eq.2.6 and the so called characteristic polynomials associated with $D_N$ or $B_N$. They are defined by

$$\phi_{D_N}(z) = \prod_{i=1}^{N} (1-t_i z)(1-t_i^{-1} z), \quad \phi_{B_N}(z) = (1-z)\prod_{i=1}^{N} (1-t_i z)(1-t_i^{-1} z)$$

where the dependence upon $t_1,\, \cdots,\, t_N$ is understood, but not written. On the other hand, it is also useful to make use of

$$\phi_{GL(M)}(z) = \prod_{i=1}^{M} (1-t_i z).$$

In the proof of the so called Littlewood's lemma the following relation was derived in ref.[7]:

$$\prod_{m=1}^{M} \frac{1}{\phi_{GL(M)}(z_m)} = \sum_{[\vec{\mu}],\, d(\vec{\mu})\leq M} \chi_{[\vec{\mu}]}^{GL(M)}(q_1, \ldots q_M) \chi_{[\vec{\mu}]}^{GL(M)}(z_1, \ldots z_M). \tag{2.7}$$

The symbols $[\vec{\mu}] = [\mu_1,\, \mu_2,\, \cdots]$ are defined so that $\mu_1 \geq \mu_2 \geq \cdots \geq 0$, with a finite number $d(\vec{\mu})$ (called the depth) being non zero. These symbols of course characterise Young diagrams. The $GL(M)$ characters are given by Schur functions which, among other ways, may be defined by

$$\chi^{GL(M)}_{[\vec{\mu}]}(q_1,\, q_2,\, \dots q_M) = \frac{\det \begin{vmatrix} q_1^{\mu_1+M-1} & q_1^{\mu_2+M-2} & \dots & q_1^{\mu_M} \\ q_2^{(\mu_1+M-1)} & q_2^{(\mu_2+M-2)} & \dots & q_2^{\mu_M} \\ \vdots & \vdots & \ddots & \vdots \\ q_M^{(\mu_1+M-1)} & q_M^{(\mu_2+M-2)} & \dots & q_M^{\mu_M} \end{vmatrix}}{\det \begin{vmatrix} q_1^{M-1} & q_1^{M-2} & \dots & 1 \\ q_2^{(M-1)} & q_2^{(M-2)} & \ddots & 1 \\ \vdots & \vdots & \vdots & \vdots \\ q_M^{(M-1)} & q_M^{(M-2)} & \dots & 1 \end{vmatrix}}. \tag{2.8}$$

[5] A priori one expects that this should be possible only at the critical dimension where $N=12$. However, as already argued in refs.[2] this is also true for general $N$.

In this formula and throughout the article, $q_j^k$ means $q_j$ to the power $k$, for any lower index $j$. The stringy character Eq.2.6 may be written as

$$\chi(q_0\,|\,q_1,\,\cdots,q_N) = \prod_{n\geq 1}\frac{1}{\phi_{D_N}(q_0^n)} = \prod_{n\geq 1}\frac{1-q_0^n}{\phi_{B_N}(q_0^n)} \equiv f_0\prod_{n\geq 1}\frac{1}{\phi_{B_N}(q_0^n)},$$

where we have defined

$$f_0 = \prod_{n\geq 1}(1-q_0^n). \tag{2.9}$$

We start with a cutoff $M$ on the mode number, and consider

$$\chi^{D_N,M}(q_0\,|\,q_1,\,\dots q_N) \equiv \prod_{\nu=1}^{N}\prod_{m=1}^{M}\frac{1}{1-q_0^m q_\nu}\frac{1}{1-(q_0^m/q_\nu)} = \prod_{n=1}^{M}\frac{1}{\phi_{D_N}(q_0^n)}.$$

The particular pairing between the terms ($q_\mu$, and $q_\mu^{-1}$) obviously come from the orthogonality condition. In order to use Eq.2.7, we first relax this condition and expand into irreducible characters of $GL(2N)$. If we introduce, in general, the $GL(P)$ stringy character

$$\chi^{GL(P),M}(q_0\,|\,q_1,\,\dots q_P) \equiv \prod_{\nu=1}^{P}\prod_{m=1}^{M}\frac{1}{1-q_0^m q_\nu},$$

we have

$$\chi^{D_N,M}(q_0\,|\,q_1,\,\dots q_N) \equiv \chi^{GL(2N),M}(q_0\,|\,q_1,\,\dots q_N,\,q_1^{-1},\,\dots q_N^{-1}),$$

$$\chi(q_0\,|\,q_1,\,\dots q_N) = \lim_{M\to\infty} \chi^{GL(2N),M}(q_0\,|\,q_1,\,\dots q_N,\,q_1^{-1},\,\dots q_N^{-1}). \qquad (2.10)$$

So we concentrate first on the $GL$ character. We only need $\chi^{GL(P),M}(q_0\,|\,q_1,\,\dots,\,q_P)$ for $P<M$, so that we may write,

$$\chi^{GL(P),M}(q_0\,|\,q_1,\,\dots,\,q_P) = \widetilde{\rho}_{M,P}\left(\chi^{GL(M),M}(q_0\,|\,q_1,\,\dots,\,q_M)\right),$$

where, using ref.[7]'s notation, $\widetilde{\rho}_{M,P}$ means that we let the last $P-M$ variables equal zero. Eq.2.7 gives

$$\chi^{GL(M),M}(q_0\,|\,q_1,\,\dots,\,q_M) = \sum_{[\vec{\mu}],\,d(\vec{\mu})\le M} \chi^{GL(M)}_{[\vec{\mu}]}(q_1,\dots,\,q_M)\chi^{GL(M)}_{[\vec{\mu}]}(q_0,\,q_0^2,\,\dots,\,q_0^M),$$

and therefore

$$\chi^{GL(P),M}(q_0\,|\,q_1,\,\dots q_P) = \sum_{[\vec{\mu}],\,d(\vec{\mu})\le M} \widetilde{\rho}_{M,P}\left(\chi^{GL(M)}_{[\vec{\mu}]}(q_1,\dots q_M)\right)\chi^{GL(M)}_{[\vec{\mu}]}(q_0,q_0^2,\dots,q_0^M).$$

It is easy to see, on the explicit expression Eq.2.8 that

$$\widetilde{\rho}_{M,P}\left(\chi^{GL(M)}_{[\vec{\mu}]}(q_1,\dots q_M)\right) = \chi^{GL(P)}_{[\vec{\mu}]}(q_1,\dots q_P), \quad \text{if } d(\vec{\mu})\le P,$$

and

$$\widetilde{\rho}_{M,P}\left(\chi^{GL(M)}_{[\vec{\mu}]}(q_1,\dots q_M)\right) = 0, \quad \text{if } d(\vec{\mu}) > P.$$

Thus we find that

$$\chi^{GL(P),M}(q_0\,|\,q_1,\,\dots q_P) = \sum_{[\vec{\mu}],\,d(\vec{\mu})\le P} \chi^{GL(P)}_{[\vec{\mu}]}(q_1,\dots q_P)\chi^{GL(M)}_{[\vec{\mu}]}(q_0,\,q_0^2,\,\dots,\,q_0^M).$$

Letting $M$ to be arbitrary large, we get

$$\chi^{GL(P)}(q_0\,|\,q_1,\,\dots q_P) = \sum_{[\vec{\mu}],\,d(\vec{\mu})\le P} \chi^{GL(P)}_{[\vec{\mu}]}(q_1,\dots q_P)\chi^{GL}_{[\vec{\mu}]}(q_0,\,q_0^2,\,\dots).$$

For strings, we are interested in $\chi^{GL(2N)}(q_0\,|\,q_1,\,\dots q_N,\,q_1^{-1}\dots,\,q_N^{-1})$, so

$$\chi(q_0\,|\,q_1,\,\dots q_N) = \sum_{[\vec{\mu}],\,d(\vec{\mu})\le 2N} \chi^{GL(2N)}_{[\vec{\mu}]}(q_1,\,\dots q_N,\,q_1^{-1},\,\dots,\,q_N^{-1})\chi^{GL}_{[\vec{\mu}]}(q_0,\,q_0^2,\,\dots), \qquad (2.11)$$

which is the formula we wanted to derive. Using Eq.2.8 it is easy to show that, for arbitrary $d>d(\vec{\mu})$,

$$\chi^{GL}_{[\vec{\mu}]}(q_0,\,q_0^2,\,\dots) = \chi^{GL(d)}_{[\vec{\mu}]}(q_0,\,q_0^2,\,\dots,\,1q_0^d)\prod_{i=1}^{d}\prod_{k=1}^{\mu_i}\frac{1}{1-q_0^{k+d-i}},$$

so that the above $M\to\infty$ limit is perfectly well defined.

## 2.3 Level expansion

The level structure is obtained by expanding the above character formulae in powers of $q$. The corresponding expanding of $\chi^{GL}_{[\vec\mu]}(q_0,\, q_0^2,\, \ldots)$ is based on the formula[6]

$$\chi^{GL}_{[\vec\mu]}(x_1, \ldots, x_n) = \sum_{\vec a} K^{[\vec\mu]}_{\vec a} \left( \prod_{i=1}^{n} x_i^{a_i} \right), \tag{2.12}$$

where $\vec a = (a_1 \cdots a_n)$ has non negative integer components, and $K^{[\vec\mu]}_{\vec a}$ is the number of standard Young tableaux of shape defined by $[\vec\mu]$ containing the numbers $1, 2, \ldots, n$ precisely $a_1, a_2, \ldots, a_n$ times respectively. Therefore $\sum_i a_i = \sum_i \mu_i$. In such a standard Young tableau, the numbers are inserted so that they are non decreasing going right on the same line, and strictly encreasing going down in the same column. Using Eq.2.11 we obtain

$$\chi(q_0 \,|\, q_1,\, \ldots,\, q_N) = \sum_L q_0^L \chi_L(q_1,\, \ldots q_N)$$

$$\chi_L(q_1,\, \ldots q_N) =$$

$$\sum_{p_1, \ldots, p_L} \delta(L - \sum_{m=1}^{L} m p_m) \sum_{[\vec\mu],\, d(\vec\mu) \le 2N} K^{[\vec\mu]}_{\vec p} \chi^{GL(2N)}_{[\vec\mu]}(q_1,\, \ldots q_N,\, q_1^{-1} \ldots,\, q_N^{-1}). \tag{2.13}$$

## 2.4 Expansion in irreducible characters of orthogonal groups

### 2.4.1 Expansion in terms of $D_N$ characters.

As shown, for instance in ref.[7], one has

$$\chi^{GL(2N)}_{[\vec\mu]}(q_1,\, \ldots q_N,\, q_1^{-1} \ldots,\, q_N^{-1}) = \sum_{\vec\kappa, \vec\lambda} \mathrm{LR}^{[\vec\mu]}_{[2\vec\kappa],[\vec\lambda]} \chi^{D_P}_{[\vec\lambda]}(q_1,\, \ldots q_N),$$

where $\mathrm{LR}^{[\vec\mu]}_{[2\vec\kappa],[\vec\lambda]}$ are the so-called Littlewood-Richardson numbers[6] which are defined provided[7] $[2\vec\kappa] \subset [\vec\mu]$, and $[\vec\lambda] \subset [\vec\mu]$. Thus $d(2\vec\kappa) \le 2N$, and $d(\vec\lambda) \le 2N$. The connection between Young tableaux and $D_N$ irreducible representation requires some care, on account of the existence of the two fundamental fermionic irreducible representations (see ref.[9]). Finally, we have the expansion

$$\chi(q_0\,|\,q_1,\,\ldots,\,q_N) = \sum_{[\vec\mu],\,d(\vec\mu)\le 2N,[\vec\lambda]} \sum_{\vec\kappa} \mathrm{LR}^{[\vec\mu]}_{[2\vec\kappa],[\vec\lambda]} \chi^{D_N}_{[\vec\lambda]}(q_1,\,\ldots,\,q_N)\chi^{GL}_{[\vec\mu]}(q_0,\,q_0^2,\,\ldots). \tag{2.14}$$

[6] The notation $[2\vec\kappa]$ means that one retains only Young tableaux with even number of lines.

[7] The notion $\subset$ used here is intuitive: $[\vec\nu] \subset [\vec\mu]$ means $\nu_i \le \mu_i$ $\forall i$.

### 2.4.2 Expansion in terms of $B_N$ characters.

Due to the stringy origin of the character, there exists an alternative expansion in terms of irreducible $B_N$ characters. Thus we now consider

$$\chi^{B_N,M}(q_0\,|\,q_1,\,\ldots q_N) \equiv \prod_{m=1}^{M}\left[\frac{1}{1-q_0^m}\prod_{\nu=1}^{N}\frac{1}{1-q_0^m q_\nu}\frac{1}{1-(q_0^m/q_\nu)}\right]$$

Obviously

$$\chi^{B_N,M}(q_0\,|\,q_1,\,\ldots q_N) \equiv \chi^{GL(2N+1),M}(q_0\,|\,q_1,\,\ldots q_N,\,1,\,q_1^{-1},\,\ldots q_N^{-1}),$$

$$\chi(q_0\,|\,q_1,\,\ldots q_N) = \prod_{\ell=1}^{\infty}(1-q_0^\ell)\lim_{M\to\infty}\chi^{B_N,M}(q_0\,|\,q_1,\,\ldots q_N).$$

One has

$$\chi^{GL(2N+1)}(q_0\,|\,q_1,\,\ldots q_N,\,1,\,q_1^{-1}\ldots,\,q_N^{-1}) = \sum_{[\vec\mu],\,d(\vec\mu)\le 2N+1}\chi^{GL(2N+1)}_{[\vec\mu]}(q_1,\,\ldots q_N,\,1,\,q_1^{-1}\ldots,\,q_N^{-1})\chi^{GL}_{[\vec\mu]}(q_0,\,q_0^2,\,\ldots).$$

So altogether

$$\chi(q_0\,|\,q_1,\,\ldots q_N) = \prod_{\ell=1}^{\infty}(1-q_0^\ell)\times \sum_{[\vec\mu],\,d(\vec\mu)\le 2N+1}\chi^{GL(2N+1)}_{[\vec\mu]}(q_1,\,\ldots q_N,\,1,\,q_1^{-1}\ldots,\,q_N^{-1})\chi^{GL}_{[\vec\mu]}(q_0,\,q_0^2,\,\ldots).$$

The expression in terms of $B_N$ characters follows from the relation:

$$\chi^{GL(2N+1)}_{[\vec\mu]}(q_1,\,\ldots q_N,\,1,\,q_1^{-1}\ldots,\,q_N^{-1}) = \sum_{\vec\kappa}\mathrm{LR}^{[\vec\mu]}_{[2\vec\kappa],[\vec\lambda]}\chi^{B_N}_{[\vec\lambda]}(q_1,\,\ldots q_N).$$

So we get

$$\chi(q_0\,|\,q_1,\,\ldots q_N) = \prod_{\ell=1}^{\infty}(1-q_0^\ell)\times \sum_{[\vec\mu],\,d(\vec\mu)\le 2N+1}\chi^{GL}_{[\vec\mu]}(q_0,\,q_0^2,\,\ldots)\sum_{\vec\kappa}\mathrm{LR}^{[\vec\mu]}_{[2\vec\kappa],[\vec\lambda]}\chi^{B_N}_{[\vec\lambda]}(q_1,\,\ldots q_N). \tag{2.15}$$

## 2.5 Folding

The actual work is not finished, however. Both expansions 2.14, 2.15 involve Young diagrams with depth $N < d \leq 2N$ which should be re-expressed by the so-called folding algorithm[7] in terms of true irreducible characters. Another difficulty is that the method just summarised leaves the positivity of the coefficients in doubt. Elaborating on these points is beyond the scope of the present article.

## 2.6 Highest weight states in string theories.

Another way to classify representations is to look for highest weight (also called dominant) states. To our knowledge this was never done in the string spectrum. Since it is not central here, this viewpoint is summarised in appendix A.

# 3 Modular properties of the closed bosonic string characters

## 3.1 Integration over zero-mode momentum

We concentrate on closed string since this is necessary for modular invariance. In order to calculate the equivalent of the functional integration over the torus, we should let the total momentum fluctuate. Thus we want to consider[8]

$$\chi_{\text{closed}}(q_0, \overline{q}_0 \,|\, \vec{v}) = \mathrm{Tr}\left\{ q_0^{L_0-1} \overline{q}_0^{\widetilde{L}_0-1} \prod_\mu e^{2i\pi v_\mu \mathcal{H}_\mu} \right\},$$

where

$$L_0 = \frac{1}{8}\sum_\mu (p^\mu)^2 + N_T, \quad \widetilde{L}_0 = \frac{1}{8}\sum_\mu (p^\mu)^2 + \widetilde{N}_T$$

At this point space-time is still uncompactified. There is of course only one zero mode momentum for left and right modes. Note that the oscillator space now span representations of $O(24)_L \times O(24)_R$, but only the diagonal subgroup is physical. We have let, as compared with earlier chiral formulae (see e.g. Eq.2.5),

$$q_\mu = e^{2i\pi v_\mu}, \text{ for } \mu \neq 0. \tag{3.1}$$

[8]We choose $p^+ = 1$. Any other choice may be re-absorbed by a rescaling of $q_0$ and $\overline{q}_0$.

Since we are dealing with characters of orthogonal groups, and $\prod_{\mu>0} q_\mu^{\mathcal{H}_\mu}$ is obtained in principle by diagonalising a generic group element, one would think a priori that the $v$'s are real. We will see that this contradicts modular invariance. This fact, which is due to the existence of a modified complex structure on the torus will lead us to slightly modify the above definition. The zero mode part of the Cartan generators 2.4 is:

$$h_\mu = (x^\mu p^{-\mu} - x^{-\mu} p^\mu).$$

So we get

$$\prod_{\mu>0} \int dp^\mu dp^{-\mu} \, (q_0 \overline{q}_0)^{(p^\mu)^2/8 + (p^{-\mu})^2/8} < p^\mu, p^{-\mu} | q_\mu^{h_\mu} | p^\mu, p^{-\mu} > .$$

Each term involves an integration in the plane. So, for each $\mu$, we have to deal with a two dimensional integral of the type

$$\int d_2 p < \vec{p} | \, (q_0 \overline{q}_0)^{\vec{p}^2/8} \, | \vec{p}^v >$$

where $\vec{p}^v$ is the vector rotated by an angle $2\pi v$. The integral is concentrated at $\vec{p} = 0$, and we get

$$\int d_2 p \delta [p_1 (1 - \cos(2\pi v)) - p_2 \sin(2\pi v)] \delta([p_2 (1 - \cos(2\pi v)) + p_2 \sin(2\pi v))])$$

$$= \frac{1}{4 \sin^2(\pi v))}$$

Thus we have, collecting[9] this with the non zero modes,

$$\chi_{\rm closed}(q_0, \, \overline{q}_0 \, | \, \vec{v}) =$$

$$q_0^{-1} \overline{q}_0^{-1} \prod_{\mu=1}^{12} \prod_{n \geq 1} \frac{1}{(1 - q_0^n q_\mu)} \frac{1}{(1 - q_0^n / q_\mu)} \frac{1}{(1 - \overline{q}_0^n q_\mu)} \frac{1}{(1 - \overline{q}_0^n / q_\mu)} \frac{1}{\sin^2 \pi v_\mu}.$$

## 3.2 Modular invariance for uncompactified target space

Recall that, in terms of standard theta functions,

$$\sin(\pi v) \prod_{n \geq 1} (1 - 2 e^{2in\pi\tau} \cos(2\pi v) + e^{4in\pi\tau}) = \frac{\theta_1(v|\tau)}{(\theta_1'(0|\tau))^{1/3}} e^{-i\pi\tau/6} (\frac{\pi}{2})^{2/3}.$$

[9]We drop irrelevent overall constants throughout.

Thus, we let, as usual,

$$q_0 = \exp(2i\pi\tau), \quad \overline{q}_0 = q_0^* = \exp(-2i\pi\tau^*).$$

Then (the factors $q_0\overline{q}_0$ cancell since $D = 26$)

$$\chi_{\rm closed}(q_0,\, q_0^* \,|\, \vec{v},) \equiv \chi_{\rm closed}(\tau \,|\, \vec{v}) = \prod_{\mu=1}^{12} \frac{(\theta_1'(0|\tau))^{1/3}}{\theta_1(v_\mu|\tau)} \left[\frac{(\theta_1'(0|\tau))^{1/3}}{\theta_1(v_\mu|\tau)}\right]^* \qquad (3.2)$$

The non trivial modular transformation corresponds to

$$\theta_1(\frac{v}{\tau}| - \frac{1}{\tau}) = -i(i\tau)^{1/2} e^{i\pi v^2/\tau} \theta_1(v|\tau).$$

An easy computation gives

$$\chi_{\rm closed}\left(-\frac{1}{\tau} \,|\, \frac{1}{\tau}\vec{v}\right) = e^{i\pi(\vec{v}^2/\tau - (\vec{v}^2/\tau)^*)} \chi_{\rm closed}(\tau \,|\, \vec{v})$$

This is not quite modular invariant. As we already mentionned, for the orthogonal group, the $\vec{v}$'s are real. However, since $\tau$ is complex, the transformed parameters $\vec{v}/\tau$ are complex. Thus we consider $\vec{v}$ to be complex in general[10]. It is easy to see that the modular invariant character is

$$\chi_{\rm modular}(\tau \,|\, \vec{v}) = e^{-2\pi(\Im\vec{v})^2/\Im(\tau)} \chi_{\rm closed}(\tau \,|\, \vec{v})$$

Indeed

$$\chi_{\rm modular}\left(-\frac{1}{\tau} \,\middle|\, \frac{\vec{v}}{\tau}\right) = \chi_{\rm modular}(\tau \,|\, \vec{v}).$$

Moreover, the additional factor is invariant under the other modular transformation $\tau \to \tau + 1$. Thus in general, for integer $a, b, c, d$ with $ad - bc = 1$,

$$\chi_{\rm modular}\left(\frac{a\tau + b}{c\tau + d} \,\middle|\, \frac{\vec{v}}{c\tau + d}\right) = \chi_{\rm modular}(\tau \,|\, \vec{v}) \qquad (3.3)$$

As usual, modular invariance holds only when the condition $L_0 = \widetilde{L}_0$ is not enforced. Note that, if the $v$'s are real, $\chi_{\rm modular} = \chi_{\rm closed}$. We will not elaborate upon this additional factor, since it does not appear for superstrings.

[10] This is natural for the oscillator modes, since we may apply a different $O(24)$ rotation on each chiral component.

## 3.3 Duality for circle compactification

Now one coordinate is on a circle of radius $R$, so that

$$L_0 = \frac{1}{2}(\frac{1}{2}p - L)^2 + \frac{1}{8}\sum_i p^i p^i + N_T$$

$$\widetilde{L}_0 = \frac{1}{2}(\frac{1}{2}p + L)^2 + \frac{1}{8}\sum_i p^i p^i + \widetilde{N}_T$$

### 3.3.1 The character

The invariance group becomes $O(23)$. One cannot put all transverse components in pairs any longer. A convenient convention is to compactify the component with number 12 and to leave out the one with number $-12$. Now the Cartan subalgebra has generators $\mathcal{H}_\mu$ for $\mu = 1, \cdots, 11$. So we consider

$$\chi^{\text{circ}}_{\text{closed}}(\tau|\,\vec{v}) = \text{Tr}\left\{ e^{2i\pi\tau(L_0-1)} e^{-2i\pi\tau^*\left(\widetilde{L}_0-1\right)} \prod_{\mu=1}^{11} e^{2i\pi v_\mu \mathcal{H}_\mu} \right\}$$

The zero mode part is

$$\int d^{(22)}p dp^{-12} < p| e^{2i\pi(\tau-\tau^*)(\sum_i p^i p^i/2 + p^{-12}p^{-12}/2)} \prod_{\mu=1}^{11} e^{2i\pi v_\mu h_\mu} |p> \times$$

$$\sum_{m,n} e^{i\pi[\tau(m/2R-nR)^2 - \tau^*(m/2R+nR)^2]},$$

which gives, up to irrelevent factors,

$$\frac{1}{\sqrt{(\tau-\tau^*)}} \prod_{\mu=1}^{11} \frac{1}{\sin^2(\pi v_\mu)} \sum_{m,n} e^{i\pi(\tau-\tau^*)(m^2/2R^2+n^2R^2)} e^{-i\pi(\tau+\tau^*)(mn)}.$$

Thus we have

$$\chi^{\text{circ}}_{\text{closed}}(\tau, R|\,\vec{v}) = \frac{1}{\sqrt{(\tau-\tau^*)}} \sum_{m,n} e^{i\pi(\tau-\tau^*)(m^2/rR^2+n^2R^2)} e^{-i\pi(\tau+\tau^*)mn} \times$$

$$\prod_{\mu=1}^{11} \frac{1}{\sin^2(\pi v_\mu)} \text{Tr} \prod_{n\geq 1} \left\{ e^{2i\pi\tau(\sum_{\mu=1}^{12} N_n^\mu)} e^{-2i\pi\tau^*(\sum_{\mu=1}^{12} \widetilde{N}_n^\mu)} \prod_{\mu=1}^{11} e^{2i\pi v_\mu\left(\mathcal{H}_n^\mu + \widetilde{\mathcal{H}}_n^\mu\right)} \right\}.$$

For the oscillator modes, the calculation is the same as before, with $v_{12}=0$. So we find

$$\chi^{\rm circ}_{\rm closed}(\tau,R|\,\vec v) = \frac{1}{\sqrt{\tau-\tau^*}}\sum_{m,n} e^{i\pi(\tau-\tau^*)(m^2/4R^2+n^2R^2)}e^{-i\pi(\tau+\tau^*)mn}\times$$

$$e^{-2\pi(\Im\vec v)^2/\Im(\tau)}\left|\frac{1}{(\theta_1'(0|\tau))^{2/3}}\right|^2\prod_{\mu=1}^{11}\frac{(\theta_1'(0|\tau))^{1/3}}{\theta_1(v_\mu|\tau)}\left[\frac{(\theta_1'(0|\tau))^{1/3}}{\theta_1(v_\mu|\tau)}\right]^*$$

### 3.3.2 Modular invariance

One has to deal with the sum over $m,n$. The discussion is similar to the case of the one-loop term for toroidally compactifed theories. It is convenient to define

$$F(R,\tau)=\frac{1}{\sqrt{\tau-\tau^*}}\left|\frac{1}{(\theta_1'(0|\tau))^{2/3}}\right|^2\sum_{m,n} e^{i\pi(\tau-\tau^*)(m^2/4R^2+n^2R^2)}e^{-i\pi(\tau+\tau^*)mn}.$$

One verifies that

$$F(R,\tau)=F(R,-\frac{1}{\tau})=F(\frac{1}{2R},\tau).$$

The first relation comes out by standard Poisson resummation technics (see e.g. ref.[3]). The second comes from the symmetry of the zero mode spectrum between Kaluza-Klein momentum and winding number. We may write the character as

$$\chi^{\rm circ}_{\rm closed}(\tau,R|\,\vec v) = F(R,\tau)e^{-2\pi(\Im\vec v)^2/\Im(\tau)}\prod_{\mu=1}^{11}\frac{(\theta_1'(0|\tau))^{1/3}}{\theta_1(v_\mu|\tau)}\left[\frac{(\theta_1'(0|\tau))^{1/3}}{\theta_1(v_\mu|\tau)}\right]^*$$

The product of the last three factors is modular invariant, since it is similar to the uncompactifed expression. Thus we see that

$$\chi^{\rm circ}_{\rm closed}(\tau,R|\,\vec v)=\chi^{\rm circ}_{\rm closed}(-1/\tau,R|\,\vec v/\tau)$$

# 4 Chiral superstring characters

From now on we of course let $N=4$. As a preparation for the discussion of type II superstring in the next section, we now study the chiral characters both in the Neveu-Schwarz, and Green-Schwarz formalism and establish their equivalence.

## 4.1 The NS-R formalism

The forthcoming calculation of the superstring character in the NS-R formulation was already performed in ref.[2]. The formulae are analogous to the bosonic ones except that the $b$ and $d$ are world sheet fermions. So we let, for $\mu > 0$,

$$\beta_r^{\mu} = \frac{1}{\sqrt{2}}(b_r^{\mu} - i b_r^{-\mu}), \quad \beta_r^{-\mu} = \frac{1}{\sqrt{2}}(b_r^{\mu} + i b_r^{-\mu})$$

$$\rho_n^{\mu} = \frac{1}{\sqrt{2}}(d_n^{\mu} - i d_n^{-\mu}), \quad \rho_n^{-\mu} = \frac{1}{\sqrt{2}}(d_n^{\mu} + i d_n^{-\mu})$$

which satisfy

$$[\beta_r^{\mu}, \beta_s^{\nu}] = \delta_{r,\,-s}\delta_{\mu,\,-\nu}, \quad [\rho_n^{\mu}, \rho_m^{\nu}] = \delta_{n,\,-m}\delta_{\mu,\,-\nu}.$$

This gives

$$\mathcal{H}_{\mu}^{(b)} = \sum_{r=1/2}^{\infty} (\beta_{-r}^{-\mu}\beta_r^{\mu} - \beta_{-r}^{\mu}\beta_r^{-\mu})$$

$$\mathcal{H}_{\mu}^{(d)} = \mathcal{H}_{(0),\mu}^{(d)} + \sum_{n=1}^{\infty}(\rho_{-n}^{-\mu}\rho_n^{\mu} - \rho_{-n}^{\mu}\rho_n^{-\mu}),$$

where $\mathcal{H}_{(0),\mu}^{(d)}$ is the zero mode contribution which we do not need to write explicitly. Moreover

$$L_0^{(b)} = \sum_{r=1/2}^{\infty}\sum_{\mu=1}^{4} r(\beta_{-r}^{-\mu}\beta_r^{\mu} + \beta_{-r}^{-\mu}\beta_r^{\mu}), \quad L_0^{(d)} = \sum_{n=1}^{\infty}\sum_{\mu=1}^{4} n(\rho_{-n}^{-\mu}\rho_n^{\mu} + \rho_{-n}^{\mu}\rho_n^{-\mu})$$

### 4.1.1 The NS sectors

In order to shorten the formulae we freely change notation from $v_\mu$ to $q_\mu$, for $\mu = 1, \ldots, 4$, with $q_\mu = \exp(2\pi i v_\mu)$, whichever is more suitable. The $b$ part of the trace is given by

$$\prod_{r=1/2}^{\infty} \mathrm{Tr}_r \left\{ \prod_{\mu=1}^{4} q_0^{r(\beta_{-r}^{-\mu}\beta_r^{\mu} + \beta_{-r}^{\mu}\beta_r^{-\mu})} q_\mu^{(\beta_{-r}^{-\mu}\beta_r^{\mu} - \beta_{-r}^{\mu}\beta_r^{-\mu})} \right\} = \prod_{r=1/2}^{\infty}\prod_{\mu=1}^{4}(1+q_0^r q_\mu)((1+q_0^r/q_\mu).$$

Next we have to handle the $G$ parity projection. For this we need to insert

$$G \equiv -(-1)^{\sum_{r=1/2}^{\infty} b_{-r}^{\mu} b_r^{\mu}} \tag{4.1}$$

into the trace. This gives

$$-\prod_{r=1/2}^{\infty}\mathrm{Tr}_r\left\{(-1)^{b^\mu_{-r}b^\mu_r}\prod_{\mu=1}^{4}q_0^{r(\beta^{-\mu}_{-r}\beta^\mu_r+\beta^\mu_{-r}\beta^{-\mu}_r)}q_\mu^{(\beta^{-\mu}_{-r}\beta^\mu_r-\beta^\mu_{-r}\beta^{-\mu}_r)}\right\}$$

$$=-\prod_{r=1/2}^{\infty}\prod_{\mu=1}^{4}\mathrm{Tr}_r\left\{(-1)^{(\beta^{-\mu}_{-r}\beta^\mu_r+\beta^\mu_{-r}\beta^{-\mu}_r)}q_0^{r(\beta^{-\mu}_{-r}\beta^\mu_r+\beta^\mu_{-r}\beta^{-\mu}_r)}q_\mu^{(\beta^{-\mu}_{-r}\beta^\mu_r-\beta^\mu_{-r}\beta^{-\mu}_r)}\right\}$$

$$=-\prod_{r=1/2}^{\infty}\prod_{\mu=1}^{4}(1-q_0^rq_\mu)(1-q_0^r/q_\mu)$$

Altogether the two G-parity projected characters are given by[11]

$$\chi_{\mathrm{NS}_\pm}(q_0\,|\vec{v})=q_0^{-1/2}\prod_{n=1}^{\infty}\prod_{\mu=1}^{4}\frac{1}{(1-q_0^nq_\mu)(1-q_0^n/q_\mu)}\times$$

$$\frac{1}{2}\left\{\prod_{r=1/2}^{\infty}\prod_{\mu=1}^{4}(1+q_0^rq_\mu)((1+q_0^r/q_\mu)\mp\prod_{r=1/2}^{\infty}\prod_{\mu=1}^{4}(1-q_0^rq_\mu)((1-q_0^r/q_\mu)\right\}\quad(4.2)$$

The physical sector corresponds to $\chi_{\mathrm{NS}_+}$. On the other hand, $\chi_{\mathrm{NS}_-}$ contains the tachyon, since its expansion begin with $q_0^{-1/2}$.

### 4.1.2 The R sectors

The oscillator part of the trace is given by

$$\prod_{m=1}^{\infty}\mathrm{Tr}_m\left\{\prod_{\mu=1}^{4}q_0^{m(\rho^{-\mu}_{-m}\rho^\mu_m+\rho^\mu_{-m}\rho^{-\mu}_m)}q_\mu^{(\rho^{-\mu}_{-m}\rho^\mu_m-\rho^\mu_{-m}\rho^{-\mu}_m)}\right\}=\prod_{m=1}^{\infty}\prod_{\mu=1}^{4}(1+q_0^mq_\mu)(1+q_0^m/q_\mu).$$

The zero mode part simply gives the irreducible characters of the two spinor representations of $O(8)=D_4$ which will be rederived below. In general, we shall denote the $O(8)$ irreducible characters by $\chi^{O(8)}_{(n_1,n_2,n_3,n_4)}$, where the four integers $n$ are such that the highest weight is of the form $\sum_{i=1}^4 n_i\vec{\lambda}_i$, where $\vec{\lambda}_i$ are the fundamental weights. So, the zero mode characters is either $\chi^{O(8)}_{(0,0,1,0)}(\vec{v})$ or $\chi^{O(8)}_{(0,0,0,1)}(\vec{v})$, depending upon the chirality projection ($-1$ or

[11] The factor in front is of course determined by the mass shell condition to be $q_0^{-1/2}$.

+1 respectively). Concerning the GSO projection, the Weyl condition for arbitrary fermion mass level is defined from the chirality operator

$$K = \Gamma^{10}(-1)^{\sum_{\mu=-4}^{4}\sum_{1}^{\infty} d^{\mu}_{-n}d^{\mu}_{n}} = \Gamma^{10}(-1)^{\sum_{\mu=1}^{4}\sum_{1}^{\infty}(\rho^{-\mu}_{-n}\rho^{\mu}_{n}+\rho^{\mu}_{-n}\rho^{-\mu}_{n})}. \qquad (4.3)$$

Thus we need

$$\prod_{m=1}^{\infty}\mathrm{Tr}_m\prod_{\mu=1}^{4}\left\{(-1)^{(\rho^{-\mu}_{-m}\rho^{\mu}_{m}+\rho^{\mu}_{-m}\rho^{-\mu}_{m})}q_0^{m(\rho^{-\mu}_{-m}\rho^{\mu}_{m}+\rho^{\mu}_{-m}\rho^{-\mu}_{m})}q_\mu^{(\rho^{-\mu}_{-m}\rho^{\mu}_{m}-\rho^{\mu}_{-m}\rho^{-\mu}_{m})}\right\}$$

$$= \prod_{m=1}^{\infty}\prod_{\mu=1}^{4}(1-q_0^m q_\mu)(1-q_0^m/q_\mu).$$

Altogether, projecting over the $\pm 1$ eigenspace of $K$, one gets

$$\chi_{\mathrm{R}_\pm}(q_0\,|\vec{v}) = \prod_{n=1}^{\infty}\prod_{\mu=1}^{4}\frac{1}{(1-q_0^n q_\mu)(1-q_0^n/q_\mu)}\times$$

$$\left\{\frac{1}{2}\left[\prod_{n=1}^{\infty}\prod_{\nu=1}^{4}(1+q_0^n q_\nu)(1+q_0^n/q_\nu)+\prod_{n=1}^{\infty}\prod_{\nu=1}^{4}(1-q_0^n q_\nu)(1-q_0^n/q_\nu)\right]\chi^{O(8)}_{8_\pm}(\vec{v})\right.$$
$$\left.+\frac{1}{2}\left[\prod_{n=1}^{\infty}\prod_{\nu=1}^{4}(1+q_0^n q_\nu)(1+q_0^n/q_\nu)-\prod_{n=1}^{\infty}\prod_{\nu=1}^{4}(1-q_0^n q_\nu)(1-q_0^n/q_\nu)\right]\chi^{O(8)}_{8_\mp}(\vec{v})\right\}. \qquad (4.4)$$

where we have let, for convenience, $\chi^{O(8)}_{8_+}(\vec{v}) = \chi^{O(8)}_{(0,0,0,1)}(\vec{v})$, and $\chi^{O(8)}_{8_-}(\vec{v}) = \chi^{O(8)}_{(0,0,1,0)}(\vec{v})$.

## 4.2 The GS formalism

Here, of course we use operators with spinor indices $a$ and $\dot{a}$, the $O(8)$ gamma matrices taking the form

$$\gamma^\mu = \begin{pmatrix} 0 & \gamma^\mu_{a\dot{a}} \\ \gamma^\mu_{\dot{b}b} & 0 \end{pmatrix}.$$

It is covenient to handle spinors indices by means of the equivalent basis made up with the Fock space of dimension $2^4 = 16$ generated by 4 fermionic oscillator modes:

$$\kappa^\mu = \frac{i}{2}(\gamma^\mu - i\gamma^{-\mu}), \quad \kappa^{\mu *} = \frac{-i}{2}(\gamma^\mu + i\gamma^{-\mu}), \quad \mu = 1,\,2,\,3,\,4,$$

which satisfy

$$[\kappa^\mu,\, \kappa^{\nu *}]_+ = \delta_{\mu,\nu},\; [\kappa^\mu,\, \kappa^\nu]_+ = [\kappa^{\mu *},\, \kappa^{\nu *}]_+ = 0.$$

We use the basis of occupation number states states noted $|\epsilon_1, \ldots \epsilon_4)$, $\epsilon_\mu = \pm 1$, so that

$$\kappa^{*\mu}\kappa^\mu|\epsilon_1, \ldots \epsilon_4) = \frac{1}{2}(1+\epsilon_\mu)|\epsilon_1, \ldots \epsilon_4) \tag{4.5}$$

The correspondence with the $a$, $\dot a$ indices is such that

$$|\epsilon_1, \ldots \epsilon_4) \Leftrightarrow \begin{cases} |a> & \text{if an even \# of } \epsilon \text{ are } > 0 \\ |\dot a> & \text{if an odd \# of } \epsilon \text{ are } > 0 \end{cases} \tag{4.6}$$

Concerning $O(8)$ chirality, if one lets

$$\gamma^9 = \gamma^{-4} \ldots \gamma^{-1}\gamma^1 \ldots \gamma^4$$

one finds

$$\gamma^9|a>=|a>, \qquad \gamma^9|\dot a>=-|\dot a>.$$

Let us recall, for definiteness, that the GS formulation has two $O(8)$ chirality sectors $\mathrm{GS}_\pm$. In $\mathrm{GS}_-$ the world sheet field is of the type $S^a_n$ and the spinor vacuum is of the form $|\dot a>$. In $\mathrm{GS}_+$ the opposite choice of chirality is made. Let us begin with the former case. Then

$$\chi_{\mathrm{GS}_-}(q_0\,|\,\vec v) = \mathrm{Tr}\left\{ q_0^{L_0} \prod_{\mu=1}^{4} q_\mu^{\mathcal{H}_\mu} \right\},$$

with

$$L_0 = \sum_\mu \sum_n \left( \zeta^{-\mu}_{-n}\zeta^\mu_n + \zeta^\mu_{-n}\zeta^{-\mu}_n + n\sum_a S^a_{-n}S^a_n \right)$$

$$\mathcal{H}_\mu = \sum_{n\geq 1} \frac{1}{n}(\zeta^{-\mu}_{-n}\zeta^\mu_n - \zeta^\mu_{-n}\zeta^{-\mu}_n) + \frac{1}{2}\sum_{m=-\infty}^{\infty}\sum_{a,b\dot c} : S^a_{-m}\kappa^{\mu *}_{a\dot c}\kappa^\mu_{\dot c b}S^b_m :$$

Choosing the field $S$ to have an udotted spinor index will correspond to the choice $K=-1$ for the Ramond sector projection, as we will verify below. After some transformation, one finds

$$\chi_{\mathrm{GS}_-}(q_0\,|\,\vec v) = \mathrm{Tr}_0\left\{ \prod_{\mu=1}^{4} q_\mu^{\frac{1}{2}\sum_{a,b\dot c} S^a_0\kappa^{\mu *}_{a\dot c}\kappa^\mu_{\dot c b}S^b_0} \right\} \times$$

$$\prod_{n\geq 1} \mathrm{Tr}_n \left\{ \prod_{\mu=1}^{4} q_\mu^{\frac{1}{n}(\zeta_{-n}^{-\mu}\zeta_n^{\mu}-\zeta_{-n}^{\mu}\zeta_n^{-\mu})} q_0^{\zeta_{-n}^{-\mu}\zeta_n^{\mu}+\zeta_{-n}^{\mu}\zeta_n^{-\mu}} q_\mu^{\sum_{a,b\dot c} S_{-n}^a \kappa_{a\dot c}^{\mu *}\kappa_{\dot c b}^{\mu} S_n^b} \left(\frac{q_0^n}{\sqrt{q_\mu}}\right)^{\sum_a S_{-n}^a S_n^a} \right\}$$

Let us first handle the zero modes. One gets

$$\mathrm{Tr}_0 \left\{ \prod_{\mu=1}^{4} q_\mu^{\frac{1}{2}\sum_{a,b\dot c} S_0^a \kappa_{a\dot c}^{\mu *}\kappa_{\dot c b}^{\mu} S_0^b} \right\} = \chi_{(1,0,0,0)}^{O(8)}(\vec v) + \chi_{(0,0,1,0)}^{O(8)}(\vec v)$$

The first term, which is the character of the vector representation is easily seen to be given by

$$\chi_{(1,0,0,0)}^{O(8)}(\vec v) = \sum_{\mu=1}^{4}(q_\mu + q_\mu^{-1}) \tag{4.7}$$

It is useful to elaborate on the second terms. Using Eqs.4.5, 4.6, one sees that

$$\chi_{(0,0,1,0)}^{O(8)}(\vec v) \equiv \sum_{\dot a} < \dot a| \prod_\mu q_\mu^{\mathcal{H}_\mu} |\dot a >= \mathrm{Tr}\left( P_- \prod_{\mu=1}^{4} q_\mu^{\kappa^{*\mu}\kappa^{\mu}-\frac{1}{2}} \right)$$

$$= \sum_{\epsilon_1,\dots\epsilon_4=\pm 1} \prod_{\mu=1}^{4} q_\mu^{\epsilon_\mu} \left\langle \epsilon_1,\dots\epsilon_4 \right| P_- \left| \epsilon_1,\dots\epsilon_4 \right\rangle$$

where $P_-$ is the chirality projector, and we get

$$\chi_{(0,0,1,0)}^{O(8)}(\vec v) = \sum_{\substack{\epsilon_1,\dots\epsilon_4=\pm 1\\ \text{odd number}=1}} \prod_{\mu=1}^{4} q_\mu^{\frac{1}{2}\epsilon_\mu} \tag{4.8}$$

For future use we note that a similar computation gives

$$\chi_{(0,0,0,1)}^{O(8)}(\vec v) = \sum_{\substack{\epsilon_1,\dots\epsilon_4=\pm 1\\ \text{even number}=1}} \prod_{\mu=1}^{4} q_\mu^{\frac{1}{2}\epsilon_\mu} \tag{4.9}$$

For the non zero modes, we use the same Fock space basis. Thus the $S$ Fock space is represented by

$$\prod_{n\geq 1} \prod_{\substack{\epsilon_1^{(n)},\epsilon_2^{(n)},\epsilon_3^{(n)},\epsilon_4^{(n)}\\ \text{evennb}=1}} \left(S_{-n}^{|\epsilon_1^{(n)},\epsilon_2^{(n)},\epsilon_3^{(n)},\epsilon_4^{(n)})}\right)^{N_{-n}^{|\epsilon_1^{(n)},\epsilon_2^{(n)},\epsilon_3^{(n)},\epsilon_4^{(n)})}} |0>$$

where

$$N_{-n}^{|\epsilon_1^{(n)},\epsilon_2^{(n)},\epsilon_3^{(n)},\epsilon_4^{(n)})} = 0, \text{ or } 1,$$

is the occupation number of the mode considered. For each such mode we find the factor $1+q_0^n\prod_\mu q_\mu^{\epsilon_\mu^{(n)}/2}$. The bosonic part is the same as before. Thus we find the expansion

$$\chi_{\mathrm{GS}_-}(q_0\,|\,\vec{v}) = \frac{\chi^{O(8)}_{(1,0,0,0)}(\vec{v}) + \chi^{O(8)}_{(0,0,1,0)}(\vec{v})}{\prod_{\mu=1}^4 \prod_{n\geq 1}(1-q_0^n q_\mu)\,(1-q_0^n/q_\mu)} \prod_{n\geq 1} \prod_{\substack{\{\epsilon_\mu^{(n)}\} \\ \text{even nb}=1}} \left(1+q_0^n \prod_\mu q_\mu^{\epsilon_\mu^{(n)}/2}\right), \tag{4.10}$$

which we will simplify below. A similar discussion for the $\mathrm{GS}_+$ sector gives

$$\chi_{\mathrm{GS}_+}(q_0\,|\,\vec{v}) = \frac{\chi^{O(8)}_{(1,0,0,0)}(\vec{v}) + \chi^{O(8)}_{(0,0,0,1)}(\vec{v})}{\prod_{\mu=1}^4 \prod_{n\geq 1}(1-q_0^n q_\mu)\,(1-q_0^n/q_\mu)} \prod_{n\geq 1} \prod_{\substack{\{\epsilon_\mu^{(n)}\} \\ \text{odd nb}=1}} \left(1+q_0^n \prod_\mu q_\mu^{\epsilon_\mu^{(n)}/2}\right), \tag{4.11}$$

## 4.3 $O(8)$ triality transformation of irreducible characters

Due to the triality invariance of the $O(8)$ Dynkin diagram, there is a one-to-one correspondence between representations, characterised by $(n_1,\dots n_4)$, when we permute $n_1$, with $n_3$ or $n_4$. The best known example is the connection between the three eight dimensional representations $8_v \leftrightarrow (1,0,0,0)$, $8_+ \leftrightarrow (0,0,0,1)$, and $8_- \leftrightarrow (0,0,1,0)$.

### 4.3.1 The transformation of variables

Here we show that the equivalence just recalled implies simple transformation laws for the irreducible characters under specific changes of variables. To start with, let us look for a change of variables, from $\vec{v}$ (or $q_\mu$) to new variables denoted $\vec{y}$ (or $\eta_\mu$ with $\eta_\mu = \exp(2\pi i y_\mu)$) such that $\chi^{O(8)}_{(0,0,0,1)}(\vec{v}) = \chi^{O(8)}_{(1,0,0,0)}(\vec{y})$. Using the explicit expressions Eqs.4.9, 4.7, one sees that this leads to the equation

$$\chi^{O(8)}_{(0,0,0,1)}(\vec{v}) = \sum_{\substack{\epsilon_1,\dots\epsilon_4=\pm 1 \\ \text{even number}=1}} \prod_{\nu=1}^4 q_\nu^{\frac{1}{2}\epsilon_\nu}$$

$$= (q_1q_2q_3q_4)^{-1/2} + (q_1q_2q_3q_4)^{1/2} + (\frac{q_1q_2}{q_3q_4})^{1/2} + (\frac{q_1q_2}{q_3q_4})^{-1/2} +$$

$$(\frac{q_1q_3}{q_2q_4})^{1/2}+(\frac{q_1q_3}{q_2q_4})^{-1/2}+(\frac{q_1q_4}{q_2q_3})^{1/2}+(\frac{q_1q_4}{q_2q_3})^{-1/2}\equiv\sum_{\mu=1}^{4}(\eta_\mu+\eta_\mu^{-1}).$$

A simple solution is given by

$$\begin{array}{rcl}\eta_1&=&(\frac{q_1q_3}{q_2q_4})^{1/2}\\ \eta_2&=&(\frac{q_2q_3}{q_1q_4})^{1/2}\\ \eta_3&=&(q_1q_2q_3q_4)^{1/2}\\ \eta_4&=&(\frac{q_3q_4}{q_1q_2})^{1/2}\end{array}\iff\begin{array}{rcl}q_1&=&(\frac{\eta_1\eta_3}{\eta_2\eta_4})^{1/2}\\ q_2&=&(\frac{\eta_2\eta_3}{\eta_1\eta_4})^{1/2}\\ q_3&=&(\eta_1\eta_2\eta_3\eta_4)^{1/2}\\ q_4&=&(\frac{\eta_3\eta_4}{\eta_1\eta_2})^{1/2}\end{array}\qquad(4.12)$$

The corresponding linear transformation between $\vec{v}$ and $\vec{y}$ is given by

$$\vec{v}=\mathcal{M}.\vec{y}\qquad\mathcal{M}=\begin{pmatrix}1/2&-1/2&1/2&-1/2\\-1/2&1/2&1/2&-1/2\\1/2&1/2&1/2&1/2\\-1/2&-1/2&1/2&1/2\end{pmatrix}\qquad(4.13)$$

It is easy to check that $\mathcal{M}^2=1$.

Similarly we look for a change of variable, from $\vec{v}$ (or $q_\mu$) to new variables $\vec{z}$ (or $\rho_\mu$ with $\rho_\mu=\exp(2\pi i z_\mu)$) such that $\chi^{O(8)}_{(0,0,1,0)}(\vec{v})=\chi^{O(8)}_{(1,0,0,0)}(\vec{z})$. Explicitly,

$$\chi^{O(8)}_{(0,0,1,0)}(\vec{v})=\sum_{\substack{\epsilon_1,\dots\epsilon_4=\pm1\\ \text{odd number}=1}}\prod_{\nu=1}^{4}q_\nu^{\frac{1}{2}\epsilon_\nu}$$

$$=(\frac{q_1}{q_2q_3q_4})^{-1/2}+(\frac{q_1}{q_2q_3q_4})^{1/2}+(\frac{q_2}{q_3q_4q_1})^{-1/2}+(\frac{q_2}{q_3q_4q_1})^{1/2}.$$
$$+(\frac{q_3}{q_4q_1q_2})^{-1/2}+(\frac{q_3}{q_4q_1q_2})^{1/2}+(\frac{q_4}{q_1q_2q_3})^{-1/2}+(\frac{q_4}{q_1q_2q_3})^{1/2}$$

Now we let

$$\chi^{O(8)}_{(0,0,1,0)}(\vec{v})=\sum_{\mu=1}^{4}(\rho_\mu+\rho_\mu^{-1}),$$

$$\begin{array}{rcl}\rho_1&=&(\frac{q_3}{q_4q_1q_2})^{1/2}\\ \rho_2&=&(\frac{q_4}{q_1q_2q_3})^{1/2}\\ \rho_3&=&(\frac{q_1}{q_2q_3q_4})^{1/2}\\ \rho_4&=&(\frac{q_2}{q_3q_4q_1})^{1/2}\end{array}\iff\begin{array}{rcl}q_1&=&(\frac{\rho_3}{\rho_4\rho_1\rho_2})^{1/2}\\ q_2&=&(\frac{\rho_4}{\rho_1\rho_2\rho_3})^{1/2}\\ q_3&=&(\frac{\rho_1}{\rho_2\rho_3\rho_4})^{1/2}\\ q_4&=&(\frac{\rho_2}{\rho_3\rho_4\rho_1})^{1/2}\end{array}\qquad(4.14)$$

The corresponding linear transformation between $\vec{v}$ and $\vec{z}$ is given by

$$\vec{v} = \mathcal{N}.\vec{z}, \qquad \mathcal{N} = \begin{pmatrix} 1/2 & 1/2 & -1/2 & 1/2 \\ 1/2 & 1/2 & 1/2 & -1/2 \\ -1/2 & 1/2 & 1/2 & 1/2 \\ 1/2 & -1/2 & 1/2 & 1/2 \end{pmatrix}, \tag{4.15}$$

and $\mathcal{N}^2 = 1$.

Clearly the definition of these matrices is not unique so far due to the symmetry properties (that is invariance under permutations of variables and/or replacements by inverses) of the characters which are set equal. This precise choice will turn out to be natural when we treat all representations.

### 4.3.2 The associated theta function relations

Define

$$\Theta_i(\vec{v}, \sqrt{q_0}) = \prod_{\mu=1}^{4} \theta_i(v_\mu, \sqrt{q_0})$$

where $\theta$ are the usual theta functions[12]

$$\begin{aligned}
\theta_1(v_\mu, \sqrt{q_0}) &= -i f_0 (q_\mu^{1/2} - q_\mu^{-1/2}) q_0^{1/8} \prod_{n=1}^{\infty} (1 - q_0^n q_\mu)(1 - q_0^n / q_\mu) \\
\theta_2(v_\mu, \sqrt{q_0}) &= f_0 (q_\mu^{1/2} + q_\mu^{-1/2}) q_0^{1/8} \prod_{n=1}^{\infty} (1 + q_0^n q_\mu)(1 + q_0^n / q_\mu) \\
\theta_3(v_\mu, \sqrt{q_0}) &= f_0 \prod_{n=1}^{\infty} (1 + q_0^{n-1/2} q_\mu)(1 + q_0^{n-1/2} / q_\mu) \\
\theta_4(v_\mu, \sqrt{q_0}) &= f_0 \prod_{n=1}^{\infty} (1 - q_0^{n-1/2} q_\mu)(1 - q_0^{n-1/2} / q_\mu),
\end{aligned} \tag{4.16}$$

where $f_0$ is given by Eq.2.9.

The basic fact concerning triality relations is that the $\Theta$ functions so defined transform linearly under the change of variables Eqs.4.12 or 4.14, as was discovered by Jacobi (see e.g.[10]). In appendix B we give a derivation of this fact inspired by the more modern discussion of ref.[11]. Moreover, we defined the matrices $\mathcal{M}$, and $\mathcal{N}$ so that the $\Theta$ functions transform with *the same matrices* as their arguments. The explicit formulae are as follows

[12] The definition of $q_0$ which is standard in string theory, is unfortunately the square of the definition of the parameter $q$ which is usual for $\theta$ functions.

**The $\vec{v} \to \vec{y}$ transformation** One has

$$\underline{\Theta}(\vec{v}, \sqrt{q_0}) = \mathcal{M}.\underline{\Theta}(\vec{y}, \sqrt{q_0}), \qquad \text{with } \underline{\Theta} = \begin{pmatrix} \Theta_1 \\ \Theta_2 \\ \Theta_3 \\ \Theta_4 \end{pmatrix}, \tag{4.17}$$

Expanding both sides in powers of $q_0$, give all the relations of the type $\chi^{O(8)}_{(n_1,n_2,n_3,n_4)}(\vec{v}) = \chi^{O(8)}_{(n_4,n_2,n_3,n_1)}(\vec{y})$ that follow from $O(8)$ triality. In particular, the lowest non trivial order in $q_0$ gives the following relations we will need later on:

$$\begin{pmatrix} \prod_\mu (q_\mu^{1/2} - q_\mu^{-1/2}) \\ \prod_\mu (q_\mu^{1/2} + q_\mu^{-1/2}) \\ \sum_\mu (q_\mu + 1/q_\mu) \\ -\sum_\mu (q_\mu + 1/q_\mu) \end{pmatrix} = \mathcal{M} \begin{pmatrix} \prod_\mu (\eta_\mu^{1/2} - \eta_\mu^{-1/2}) \\ \prod_\mu (\eta_\mu^{1/2} + \eta_\mu^{-1/2}) \\ \sum_\mu (\eta_\mu + 1/\eta_\mu) \\ -\sum_\mu (\eta_\mu + 1/\eta_\mu) \end{pmatrix}, \tag{4.18}$$

which, of course, includes our original starting point $\chi^{O(8)}_{(0,0,0,1)}(\vec{v}) = \chi^{O(8)}_{(1,0,0,0)}(\vec{y})$. At the next level, one finds the relations

$$\chi^{O(8)}_{(2,0,0,0)}(\vec{x}) = \chi^{O(8)}_{(0,0,0,2)}(\vec{v}), \qquad \chi^{O(8)}_{(0,1,0,0)}(\vec{x}) = \chi^{O(8)}_{(0,1,0,0)}(\vec{v}).$$

The latter means, for instance,

$$\sum_{\mu<\nu} (\eta_\mu + 1/\eta_\mu)(\eta_\nu + 1/\eta_\nu) = \sum_{\mu<\nu} (q_\mu + 1/q_\mu)(q_\nu + 1/q_\nu).$$

**The $\vec{v} \to \vec{z}$ transformations.** One has

$$\underline{\Theta}(\vec{v}, \sqrt{q_0}) = \mathcal{N}.\underline{\Theta}(\vec{z}, \sqrt{q_0}), \tag{4.19}$$

and to lowest order,

$$\begin{pmatrix} \prod_\mu (q_\mu^{1/2} - q_\mu^{-1/2}) \\ \prod_\mu (q_\mu^{1/2} + q_\mu^{-1/2}) \\ \sum_\mu (q_\mu + 1/q_\mu) \\ -\sum_\mu (q_\mu + 1/q_\mu) \end{pmatrix} = \mathcal{N} \begin{pmatrix} \prod_\mu (\rho_\mu^{1/2} - \rho_\mu^{-1/2}) \\ \prod_\mu (\rho_\mu^{1/2} + \rho_\mu^{-1/2}) \\ \sum_\mu (\rho_\mu + 1/\rho_\mu) \\ -\sum_\mu (\rho_\mu + 1/\rho_\mu) \end{pmatrix} \tag{4.20}$$

Expanding both sides in powers of $q_0$, now give all the relations of the type $\chi^{O(8)}_{(n_1,n_2,n_3,n_4)}(\vec{v}) = \chi^{O(8)}_{(n_3,n_2,n_1,n_4)}(\vec{z})$ that follow from $O(8)$ triality. Note that

$$\mathcal{N} = \mathcal{U}\mathcal{M}\mathcal{U}, \qquad \mathcal{U} = \begin{pmatrix} -1 & 0 & 0 & 0 \\ 0 & 1 & 0 & 0 \\ 0 & 0 & 1 & 0 \\ 0 & 0 & 0 & 1 \end{pmatrix}.$$

It is easy to check that $(\mathcal{MU})^3 = 1$. Thus we do have a triality relationship. In terms of the $q$, $\eta$ and $\rho$ variables, $\mathcal{U}$ corresponds to

$$\left\{ \begin{array}{l} q_1 \to 1/q_1 \\ q_i \to q_i, \quad i \neq 1 \end{array} \right. \iff \left\{ \begin{array}{l} \eta_1 \to 1/\rho_1 \\ \eta_i \to \rho_i, \quad i \neq 1 \end{array} \right. \tag{4.21}$$

As far as the characters are concerned, this transformation corresponds to the relations of the type $\chi^{O(8)}_{(n_1,n_2,n_3,n_4)}(\vec{v}) = \chi^{O(8)}_{(n_1,n_2,n_4,n_3)}(\mathcal{U}.\vec{v})$. Moreover, the action of $\mathcal{MU}$ is given by $\chi^{O(8)}_{(n_1,n_2,n_3,n_4)}(\vec{v}) = \chi^{O(8)}_{(n_4,n_2,n_1,n_3)}(\mathcal{MU}.\vec{v})$. It is a circular permutation of the three numbers $n_1, n_3, n_4$, whose cube is clearly the identity.

## 4.4 Novel insights on light-cone superstring representations

### 4.4.1 The equality between GS and NS-R characters

**The GS character** We discuss the case of GS$_-$ for definiteness. The variables $\vec{y}$ (or $\eta_\mu$) arise naturally in the expression Eq.4.10 of $\chi_{\mathrm{GS}_-}$, and we get

$$\chi_{\mathrm{GS}_-}(q_0 \,|\, \vec{v}) = \Big[\chi_{(1,0,0,0)}(\vec{v}) + \chi_{(0,0,1,0)}(\vec{v})\Big] \frac{\prod_\mu (q_\mu^{1/2} - q_\mu^{-1/2})}{\prod_\mu (\eta_\mu^{1/2} + \eta_\mu^{-1/2})} \frac{\Theta_2(\vec{y}, \sqrt{q_0})}{\Theta_1(\vec{v}, \sqrt{q_0})} \tag{4.22}$$

The prefactor is simplified by using the relation

$$\chi^{O(8)}_{(1,0,0,0)}(\vec{v}) + \chi^{O(8)}_{(0,0,1,0)}(\vec{v}) = \prod_\mu (\eta_\mu^{1/2} + \eta_\mu^{-1/2}).$$

which follow from Eq.4.18. So we find

$$\chi_{\mathrm{GS}_-}(q_0 \,|\, \vec{v}) = \prod_\mu (q_\mu^{1/2} - q_\mu^{-1/2}) \frac{\Theta_2(\vec{y}, \sqrt{q_0})}{\Theta_1(\vec{v}, \sqrt{q_0})} \tag{4.23}$$

In order to establish connection with the NS-R expression, we make use of Eq.4.17. Since $\mathcal{M}$ has square one, it gives

$$\Theta_2(\vec{y}, \sqrt{q_0}) = \frac{1}{2}\left(-\Theta_1(\vec{v}, \sqrt{q_0}) + \Theta_2(\vec{v}, \sqrt{q_0}) + \Theta_3(\vec{v}, \sqrt{q_0}) - \Theta_4(\vec{v}, \sqrt{q_0})\right).$$

Therefore

$$\chi_{\mathrm{GS}_-}(q_0 \,|\, \vec{v}) =$$

$$\frac{\prod_\mu (q_\mu^{1/2} - q_\mu^{-1/2})}{2\Theta_1(\vec{v}, \sqrt{q_0})} \left[\Theta_2(\vec{v}, \sqrt{q_0}) - \Theta_1(\vec{v}, \sqrt{q_0}) + \Theta_3(\vec{v}, \sqrt{q_0}) - \Theta_4(\vec{v}, \sqrt{q_0})\right] \tag{4.24}$$

**The NSR character** Let us define $\chi_{\mathrm{NSR}_-} = \chi_{\mathrm{NS}_+} + \chi_{\mathrm{R}_-}$. Using Eqs.4.2 and 4.4 it may be written as

$$\chi_{\mathrm{NSR}_-}(q_0 \,|\, \vec{v}) = \frac{\prod_\mu (q_\mu^{1/2} - q_\mu^{-1/2})}{\Theta_1(\vec{v}, \sqrt{q_0})} \frac{1}{2} \Big\{ \Theta_3(\vec{v}, \sqrt{q_0}) - \Theta_4(\vec{v}, \sqrt{q_0}) +$$

$$\Theta_1(\vec{v}, \sqrt{q_0}) \frac{(\chi_{(0,0,1,0)}(\vec{v}) - \chi_{(0,0,0,1)}(\vec{v}))}{\prod_\mu (q_\mu^{1/2} - q_\mu^{-1/2})}$$

$$\Theta_2(\vec{v}, \sqrt{q_0}) \frac{(\chi_{(0,0,1,0)}(\vec{v}) + \chi_{(0,0,0,1)}(\vec{v}))}{\prod_\mu (q_\mu^{1/2} + q_\mu^{-1/2})} \Big\}$$

Since, according to Eq.4.18,

$$\frac{(\chi^{O(8)}_{(0,0,1,0)}(\vec{v}) \pm \chi^{O(8)}_{(0,0,0,1)}(\vec{v}))}{\prod_\mu (q_\mu^{1/2} \pm q_\mu^{-1/2})} = \pm 1,$$

we verify that $\chi_{\mathrm{GS}_-}(q_0 \,|\, \vec{v}) = \chi_{\mathrm{NSR}_-}(q_0 \,|\, \vec{v})$ as expected.

### 4.4.2 Separating the NS and R sectors in the GS formalism

Obviously we may write

$$\chi_{\mathrm{GS}_-}(q_0 \,|\, \vec{v}) =$$

$$\frac{\prod_\mu (q_\mu^{1/2} - q_\mu^{-1/2})}{\Theta_1(\vec{v}, \sqrt{q_0})} \left( \frac{1}{2} \left[\Theta_2(\vec{y}, \sqrt{q_0}) + \Theta_1(\vec{y}, \sqrt{q_0})\right] + \frac{1}{2} \left[\Theta_2(\vec{y}, \sqrt{q_0}) - \Theta_1(\vec{y}, \sqrt{q_0})\right] \right) \tag{4.25}$$

Let us show that this corresponds to the separation between NS and R sectors. Indeed, according to Eq.4.17, one has

$$\Theta_1(\vec{y}, \sqrt{q_0}) + \Theta_2(\vec{y}, \sqrt{q_0}) = \Theta_3(\vec{v}, \sqrt{q_0}) - \Theta_4(\vec{v}, \sqrt{q_0})$$

$$\Theta_1(\vec{y}, \sqrt{q_0}) - \Theta_2(\vec{y}, \sqrt{q_0}) = \Theta_1(\vec{v}, \sqrt{q_0}) - \Theta_2(\vec{v}, \sqrt{q_0})$$

So indeed,

$$\chi_{\mathrm{NS}_+}(q_0 \,|\, \vec{v}) \equiv \frac{\prod_\mu (q_\mu^{1/2} - q_\mu^{-1/2})}{\Theta_1(\vec{v}, \sqrt{q_0})} \frac{1}{2} \left[\Theta_3(\vec{v}, \sqrt{q_0}) - \Theta_4(\vec{v}, \sqrt{q_0})\right] =$$

$$\frac{\prod_\mu (q_\mu^{1/2} - q_\mu^{-1/2})}{\Theta_1(\vec{v}, \sqrt{q_0})} \frac{1}{2} \left[\Theta_1(\vec{y}, \sqrt{q_0}) + \Theta_2(\vec{y}, \sqrt{q_0})\right]$$

$$\chi_{\mathrm{R}_-}(q_0 \,|\, \vec{v}) \equiv \frac{\prod_\mu (q_\mu^{1/2} - q_\mu^{-1/2})}{\Theta_1(\vec{v}, \sqrt{q_0})} \frac{1}{2} \left[\Theta_2(\vec{v}, \sqrt{q_0}) - \Theta_1(\vec{v}, \sqrt{q_0})\right] =$$

$$\frac{\prod_\mu (q_\mu^{1/2} - q_\mu^{-1/2})}{\Theta_1(\vec{v}, \sqrt{q_0})} \frac{1}{2} \left[\Theta_2(\vec{y}, \sqrt{q_0}) - \Theta_1(\vec{y}, \sqrt{q_0})\right]$$

### 4.4.3 Projecting out the NS and R sectors in the GS formalism.

Let us return to the $\Theta_2$ term. Its expansion is

$$\Theta_2(\vec{y}, \sqrt{q_0}) = \left[\chi_{(1,0,0,0)}(\vec{v}) + \chi_{(0,0,1,0)}(\vec{v})\right] \frac{\Theta_2(\vec{y}, \sqrt{q_0})}{\prod_\mu (\eta_\mu^{1/2} + \eta_\mu^{-1/2})}$$

$$= f_0^4 q_0^{1/2} \left[\chi_{(1,0,0,0)}(\vec{v}) + \chi_{(0,0,1,0)}(\vec{v})\right] \prod_\mu \prod_{n=1}^{\infty} (1 + q_0^n \eta_\mu)(1 + q_0^n / \eta_\mu).$$

$$= f_0^4 q_0^{1/2} \left[\chi_{(1,0,0,0)}(\vec{v}) + \chi_{(0,0,0,1)}(\vec{v})\right] \prod_n \prod_{\substack{\epsilon_1^n, \epsilon_2^n, \epsilon_3^n, \epsilon_4^n \\ \text{even nb=1}}} \left(1 + q_0^n \prod_\mu q_\mu^{\epsilon_\mu^n/2}\right)$$

This corresponds to GR operators $S^a$ acting on vacua $|i>$ and $|\dot{a}>$, as we already know. Next consider $\Theta_1$. We have

$$\Theta_1(\vec{y}, \sqrt{q_0}) = f_0^4 q_0^{1/2} \prod_\mu (\eta_\mu^{1/2} - \eta_\mu^{-1/2}) \prod_\mu \prod_{n=1}^{\infty} (1 - q_0^n \eta_\mu)(1 - q_0^n / \eta_\mu).$$

$$= f_0^4 q_0^{1/2} \prod_\mu (\eta_\mu^{1/2} - \eta_\mu^{-1/2}) \prod_n \prod_{\substack{\epsilon_1^n, \epsilon_2^n, \epsilon_3^n, \epsilon_4^n \\ \text{even nb=1}}} \left(1 - q_0^n \prod_\mu q_\mu^{\epsilon_\mu^n/2}\right)$$

Consider first the zero mode contribution. One has

$$\prod_\mu (\eta_\mu^{1/2} - \eta_\mu^{-1/2}) = \sum_\mu (q_\mu + q_\mu^{-1}) - \frac{1}{2} \left(\prod_\mu (q_\mu^{1/2} + q_\mu^{-1/2}) - \prod_\mu (q_\mu^{1/2} - q_\mu^{-1/2})\right)$$

$$= \chi^{O(8)}_{(1,0,0,0)}(\vec{v}) - \chi^{O(8)}_{(0,0,1,0)}(\vec{v})$$

Let us return to the operator realisation. Define the zero mode operator $(-1)^{N_F}$, by

$$(-1)^{N_F} |i> = |i>, \qquad (-1)^{N_F} |\dot{a}> = -|\dot{a}>,$$

so that

$$(-1)^{N_F} S_0^a = -S_0^a (-1)^{N_F}$$

Then the above zero mode contribution is the trace with $(-1)^{N_F}$ inserted. Including the oscillator modes, we find

$$q_0^4 q_0^{1/2} \Theta_1(\vec{y}, \sqrt{q_0}) = \mathrm{Tr}_0 \left\{ \prod_{\mu=1}^{4} (-1)^{N_F} q_\mu^{\frac{1}{2}\sum_{a,b\dot{c}} S_0^a \kappa_{a\dot{c}}^{\mu *} \kappa_{\dot{c}b}^{\mu} S_0^b} \right\} \times$$

$$\prod_{n\geq 1} \mathrm{Tr}_n \left\{ \prod_{\mu=1}^{4} q_\mu^{\sum_{a,b,\dot{c}} S_{-n}^a \kappa_{a\dot{c}}^{\mu *} \kappa_{\dot{c}b}^{\mu} S_n^b} \left( \frac{-q_0^n}{\sqrt{q_\mu}} \right)^{\sum_a S_{-n}^a S_n^a} \right\}$$

So we have a sort of parity operator in the GS sector. As far as world sheet properties are concerned, it is a parity operator for a field with integer modes, so that it is analogous to the chirality operator of the R sector (see Eq.4.3) which we denoted $K$. Thus we call[13] it $K_S^{(-)}$. It is given by

$$K_S^{(-)} = (-1)^{N_F} (-1)^{\sum_{n\geq 1} \sum_a S_{-n}^a S_n^a} \tag{4.26}$$

Of course the space-time properties of $K_S$ are very different from those of the chirality $K$. It separates fermions and bosons since $S$ is a spinor. Since, we have found that

$$[\Theta_2(\vec{y}, \sqrt{q_0}) \pm \Theta_1(\vec{y}, \sqrt{q_0})] =$$

$$f_0^4 q_0^{1/2} \mathrm{Tr} \left\{ \left( 1 \pm (-1)^{N_F} \prod_{n\geq 1} (-1)^{\sum_a S_{-n}^a S_n^a} \right) \prod_{\mu=1}^{4} q_\mu^{\frac{1}{2}\sum_{a,b\dot{c}} S_0^a \kappa_{a\dot{c}}^{\mu *} \kappa_{\dot{c}b}^{\mu} S_0^b} \times \right.$$

$$\left. \prod_{n\geq 1} \prod_{\mu=1}^{4} q_\mu^{\sum_{a,b,\dot{c}} S_n^a \kappa_{a\dot{c}}^{\mu *} \kappa_{\dot{c}b}^{\mu} S_n^b} \left( \frac{q_0^n}{\sqrt{q_\mu}} \right)^{\sum_a S_{-n}^a S_n^a} \right\}, \tag{4.27}$$

we see that the $NS_+$ sector has positive $K_S^{(-)}$-parity, and the $R_-$ sector negative $K_S^{(-)}$-parity.

### 4.4.4 The third formalism

The $GS_-$ formalism is obtained from a mixed representation using $\eta$ in the numerator and $q$ in the denominator. Let us try a third representation that uses $\rho$ and $q$ similarly to describe *the same chirality sector* $GS_-$.

[13]The superscript $(-)$ is a reminder of the $GS_-$ sector we are presently discussing.

**The $\mathrm{NS}_+$ sector:** We use the fact that

$$\Theta_2(\vec{y}\sqrt{q_0})+\Theta_1(\vec{y},\sqrt{q_0})=\Theta_3(\vec{v},\sqrt{q_0})-\Theta_4(\vec{v},\sqrt{q_0})=\Theta_2(\vec{z},\sqrt{q_0})-\Theta_1(\vec{z},\sqrt{q_0})$$

So we have

$$\chi_{\mathrm{NS}_+}(q_0\,|\,\vec{v})\equiv\frac{\prod_\mu(q_\mu^{1/2}-q_\mu^{-1/2})}{\Theta_1(\vec{v},\sqrt{q_0})}\frac{1}{2}\left[\Theta_2(\vec{z},\sqrt{q_0})-\Theta_1(\vec{z},\sqrt{q_0})\right]$$

What is the oscillator interpretation? Obviously, for the $\Theta_2$ term it is similar to the usual GS term, albeit with opposite chirality. So, for it we write

$$\Theta_2(\vec{z},\sqrt{q_0})=\left[\chi_{(1,0,0,0)}(\vec{v})+\chi_{(0,0,0,1)}(\vec{v})\right]\frac{\Theta_2(\vec{z},\sqrt{q_0})}{\prod_\mu(\rho_\mu^{1/2}+\rho_\mu^{-1/2})}$$

$$=f_0^4q_0^{1/2}\left[\chi_{(1,0,0,0)}(\vec{v})+\chi_{(0,0,0,1)}(\vec{v})\right]\prod_\mu\prod_{n=1}^{\infty}(1+q_0^n\rho_\mu)(1+q_0^n/\rho_\mu).$$

$$=f_0^4q_0^{1/2}\left[\chi_{(1,0,0,0)}(\vec{v})+\chi_{(0,0,0,1)}(\vec{v})\right]\prod_n\prod_{\substack{\epsilon_1^n,\epsilon_2^n,\epsilon_3^n,\epsilon_4^n\\ \text{odd nb}=1}}\left(1+q_0^n\prod_\mu q_\mu^{\epsilon_\mu^n/2}\right)$$

This corresponds to GR operators $S^{\dot{a}}$ acting on vacuua $|i>$ and $|a>$, that is, to the choice of chirality opposite to the one of Green Schwarz in $\mathrm{GS}_-$. Next consider $\Theta_1$. We have

$$\Theta_1(\vec{z},\sqrt{q_0})=f_0^4q_0^{1/2}\prod_\mu(\rho_\mu^{1/2}-\rho_\mu^{-1/2})\prod_\mu\prod_{n=1}^{\infty}(1-q_0^n\rho_\mu)(1-q_0^n/\rho_\mu).$$

$$=f_0^4q_0^{1/2}\prod_\mu(\rho_\mu^{1/2}-\rho_\mu^{-1/2})\prod_n\prod_{\substack{\epsilon_1^n,\epsilon_2^n,\epsilon_3^n,\epsilon_4^n\\ \text{odd nb}=1}}\left(1-q_0^n\prod_\mu q_\mu^{\epsilon_\mu^n/2}\right)$$

Consider first the zero mode contribution. One has

$$\prod_\mu(\rho_\mu^{1/2}-\rho_\mu^{-1/2})=\frac{1}{2}\left(\prod_\mu(q_\mu^{1/2}+q_\mu^{-1/2})+\prod_\mu(q_\mu^{1/2}-q_\mu^{-1/2})\right)-\sum_\mu(q_\mu+q_\mu^{-1})$$

$$=-\chi^{O(8)}_{(1,0,0,0)}(\vec{v})+\chi^{O(8)}_{(0,0,0,1)}(\vec{v})$$

Let us return to the operator realisation. By a reasoning similar to the above, it is easy to see that

$$\left[\Theta_2(\vec{z},\sqrt{q_0})-\Theta_1(\vec{z},\sqrt{q_0})\right]=$$

$$f_0^4 q_0^{1/2} \mathrm{Tr} \left\{ \frac{1}{2} \left( 1 + (-1)^{N_F} \prod_{n \geq 1} (-1)^{\sum_{\dot a} S^{\dot a}_{-n} S^{\dot a}_n} \right) \prod_{\mu=1}^{4} q_\mu^{\frac{1}{2} \sum_{\dot a, \dot b c} S^{\dot a}_0 \kappa^{\mu *}_{\dot a c} \kappa^{\mu}_{c \dot b} S^{\dot b}_0} \times \right.$$

$$\left. \prod_{n \geq 1} \prod_{\mu=1}^{4} q_\mu^{\sum_{\dot a, \dot b c} S^{\dot a}_{-n} \kappa^{\mu *}_{\dot a c} \kappa^{\mu}_{c \dot b} S^{\dot b}_n} \left( \frac{q_0^n}{\sqrt{q_\mu}} \right)^{\sum_{\dot a} S^{\dot a}_{-n} S^{\dot a}_n} \right\}. \tag{4.28}$$

We see that another parity operator appears. We define it as

$$K_S^{(+)} = -(-1)^{N_F} (-1)^{\sum_{n \geq 1} \sum_{\dot a} S^{\dot a}_{-n} S^{\dot a}_n} \tag{4.29}$$

Thus in the third description, the $\mathrm{NS}_+$ sector is decribed by the $K_S^{(+)}$ negative-parity sector. Let us explicit verify that we get the same spectrum as in the usual formalism for lowest levels. For the vacuum, we find, according to Eq.4.20,

$$\frac{1}{2} q_0^{1/2} \left( \prod_\mu (\rho_\mu^{1/2} + \rho_\mu^{-1/2}) - \prod_\mu (\rho_\mu^{1/2} - \rho_\mu^{-1/2}) \right)$$

$$= q_0^{1/2} \sum_\mu (q_\mu + 1/q_\mu)$$

which is the correct lowest (vector) state of NS, ie $b^\mu_{-1/2}|0>$. The factor $q_0^{1/2}$ is cancelled by the $\Theta_1$ in the denominator. The next level is given by $S^{\dot a}_{-1}|a>$, that is 64 states. They may be re-arranged in the states

$$\gamma^\mu_{a\dot a} S^{\dot a}_{-1}|a>, \gamma^{\mu\nu\rho}_{a\dot a} S^{\dot a}_{-1}|a>$$

For the usual description, we have the corresponding states $b^\mu_{-3/2}|0>$, and $b^\mu_{-1/2} b^\nu_{-1/2} b^\rho_{-1/2}|0>$ with the same physical content.

What would happen without projection? Then the vacuum would be $|\alpha> + |a>$, a mixture of fermionic and bosonic states. See section 4.5 below.

**The $\mathrm{R}_-$ sector** We now use Eqs.4.20 to write

$$\Theta_2(\vec y \sqrt{q_0}) - \Theta_1(\vec y, \sqrt{q_0}) = \Theta_2(\vec v, \sqrt{q_0}) - \Theta_1(\vec v, \sqrt{q_0}) = \Theta_4(\vec z, \sqrt{q_0}) - \Theta_3(\vec z, \sqrt{q_0})$$

So we have

$$\chi_{\mathrm{R}_-}(\vec v, \sqrt{q_0}) \equiv \frac{\prod_\mu (q_\mu^{1/2} - q_\mu^{-1/2})}{\Theta_1(\vec v, \sqrt{q_0})} \frac{1}{2} \left[ \Theta_3(\vec z, \sqrt{q_0}) - \Theta_4(\vec z, \sqrt{q_0}) \right]$$

What is the oscillator interpretation? Consider

$$\Theta_3(\vec{z},\sqrt{q_0}) = q_0^8 \prod_\mu \prod_{n=1}^\infty (1+q_0^{n-1/2}\rho_\mu)(1+q_0^{n-1/2}/\rho_\mu)$$

This corresponds to a world sheet field $T_r^{\dot{a}}$ where $r$ is a half integer mode number. Indeed, we may write

$$\prod_{\mu=1}^4 \prod_{n=1}^\infty (1+q_0^{n-1/2}\rho_\mu)(1+q_0^{n-1/2}/\rho_\mu) = \prod_{r\geq 1/2} \prod_{\substack{\epsilon_1^{(r)},\epsilon_2^{(r)},\epsilon_3^{(r)},\epsilon_4^{(r)} \\ \text{odd nb}=1}} \left(1+q_0^r \prod_\mu q_\mu^{\epsilon_\mu^{(r)}/2}\right)$$

Thus the $T$ Fock space is represented by

$$\prod_{r\geq 1/2} \prod_{\substack{\epsilon_1^{(r)},\epsilon_2^{(r)},\epsilon_3^{(r)},\epsilon_4^{(r)} \\ \text{evennb}=1}} \left(T_{-r}^{|\epsilon_1^{(r)},\epsilon_2^{(r)},\epsilon_3^{(r)},\epsilon_4^{(r)})}\right)^{N_{-r}^{|\epsilon_1^{(r)},\epsilon_2^{(r)},\epsilon_3^{(r)},\epsilon_4^{(r)})}} |0>$$

where $N_{-r}^{|\epsilon_1^{(r)},\epsilon_2^{(r)},\epsilon_3^{(r)},\epsilon_4^{(r)})}$ is zero or one. Since we consider $\Theta_3-\Theta_4$, we should introduce a parity. Since $T$ has half integer modes it is similar to the $G$ parity of the NS sector, and we call it $G_T^{(+)}$. The operator is given by

$$G_T^{(+)} = -(-1)^{\sum_{r\geq 1/2}\sum_{\dot{a}} T_{-r}^{\dot{a}} T_r^{\dot{a}}} \tag{4.30}$$

We are retaining the sector with positive $G_T^{(+)}$ parity only. Thus the vaccuum is $T_{-1/2}^{\dot{a}}|0>$, which corresponds to the NS vacuum $|\dot{a}>$. The corresponding contribution is

$$q_0^{1/2}\sum_\mu(\rho_\mu+1/\rho_\mu) = q_0^{1/2}\chi_{(1,0,0,0))}(\vec{z}) = q_0^{1/2}\chi_{(0,0,0,1))}(\vec{v})$$

Again, the $q_0^{1/2}$ is cancelled by the $\Theta_1$ in the denominator. The next level is $T_{-3/2}^{\dot{a}}|0>$ and $T_{-1/2}^{\dot{a}}T_{-1/2}^{\dot{b}}T_{-1/2}^{\dot{c}}|0>$, that is 64 states. What are the corresponding states of GS? One has $S_{-1}^a|\alpha>$, which also gives 64 states. The state in the representation $(0,0,1,0)$ is given by $\gamma_{\dot{a}a}^\alpha S_{-1}^a|\alpha>$. The rest has dimension 56. In the R approach, the corresponding states are $d_{-1}^\mu|a>$ which also gives 64 states, the $(0,0,1,0)$ irrep being again $\gamma_{\dot{a}a}^\mu d_{-1}^\mu|a>$. So all is of course consistent.

What would happen if we did not project out: Then we have the constant term of the $\Theta_3$ expansion, which is a singlet, but there is still the $q_0^{-1/2}$ due to the denominator. Thus there is still the bosonic tachyon. See next section.

## 4.5 The general picture

Let us keep all sectors in every formulations to show that they are completely equivalent. The correspondence between characters may be summarized as follows[14]

| Sector | Theta(v) | parity | Theta(y) | parity | Theta(z) | parity |
|---|---|---|---|---|---|---|
| $NS_-$ | $\Theta_3+\Theta_4$ | $G=-$ | $\Theta_3+\Theta_4$ | $G_T^{(-)}=-$ | $\Theta_3+\Theta_4$ | $G_T^{(+)}=-$ |
| $NS_+$ | $\Theta_3-\Theta_4$ | $G=+$ | $\Theta_2+\Theta_1$ | $K_S^{(-)}=+$ | $\Theta_2-\Theta_1$ | $K_S^{(+)}=-$ |
| $R_+$ | $\Theta_2+\Theta_1$ | $K=+$ | $\Theta_3-\Theta_4$ | $G_T^{(-)}=+$ | $\Theta_2+\Theta_1$ | $K_S^{(+)}=+$ |
| $R_-$ | $\Theta_2-\Theta_1$ | $K=-$ | $\Theta_2-\Theta_1$ | $K_S^{(-)}=-$ | $\Theta_3-\Theta_4$ | $G_T^{(+)}=+$ |

Note that the tachyon sector appears in every formulations including the GS one. In $NRS_- = NS_+ \oplus R_-$, the usual GS formulation corresponds to using $S$ fields only, since then both sectors $K_S^{(-)} = \pm$ are physical. In this formulation the GSO projection just corresponds to discarding the $T$ fields. However, this sector still exists. It is useful if one wants to use the third type of formulation for the other GSO projection $NRS_+ = NS_+ \oplus R_+$. Then the physical sectors have $K_S^{(-)} = +$ in the $S$ sector, and $G_T^{(-)} = +$ in the $T$ sector. The two GSO projected sectors are described as follows

| Sector | Theta(v) | parity | Theta(y) | parity | Theta(z) | parity |
|---|---|---|---|---|---|---|
| $NS_+$ | $\Theta_3-\Theta_4$ | $G=+$ | $\Theta_2+\Theta_1$ | $K_S^{(-)}=+$ | $\Theta_2-\Theta_1$ | $K_S^{(+)}=-$ |
| $R_+$ | $\Theta_2+\Theta_1$ | $K=+$ | $\Theta_3-\Theta_4$ | $G_T^{(-)}=+$ | $\Theta_2+\Theta_1$ | $K_S^{(+)}=+$ |
| | Neveu Schwarz Ramond | | Third formulation | | Green Schwarz | |

The $GSO_+$ sector $NSR_+$

| Sector | Theta(v) | parity | Theta(y) | parity | Theta(z) | parity |
|---|---|---|---|---|---|---|
| $NS_+$ | $\Theta_3-\Theta_4$ | $G=+$ | $\Theta_2+\Theta_1$ | $K_S^{(-)}=+$ | $\Theta_2-\Theta_1$ | $K_S^{(+)}=-$ |
| $R_-$ | $\Theta_2-\Theta_1$ | $K=-$ | $\Theta_2-\Theta_1$ | $K_S^{(-)}=-$ | $\Theta_3-\Theta_4$ | $G_T^{(+)}=+$ |
| | Neveu Schwarz Ramond | | Green Schwarz | | Third formulation | |

The $GSO_-$ sector $NSR_-$

[14]up to a common factor 1/2 which we remove for brevity,

It is interesting to note that the character corresponding to the complete NSR spectrum, that is *without any projection*, is given by the same $\Theta_2+\Theta_3$ function in terms of $\vec{v}$ or $\vec{y}$ or $\vec{z}$. In appendix C, we show how the new $T^a$ and $T^{\dot{a}}$ fields we have introduced may be constructed using the same bosonic fields as the usual GS and NSR free fields.

## 4.6 The impact of supersymmetry: Factorising the long SUSY multiplet character for massive levels.

Consider $\chi_{\mathrm{GS}_-}$ for definiteness. Expand it up to the first massive level. One may either use Eqs.4.10, and 4.16, or return to the original operator formulation, to derive

$$\chi_{\mathrm{GS}_-}(q_0\,|\,\vec{v}) \sim \chi^{(O(8)}_{(0,0,1,0)}(\vec{v}) + \chi^{(O(8)}_{(1,0,0,0)}(\vec{v}) + q_0\chi^{\mathrm{long}}(\vec{v}) + \cdots$$

where

$$\chi^{\mathrm{long}}(\vec{v}) = (\chi^{(O(8)}_{(1,0,0,0)}(\vec{v}) + \chi^{(O(8)}_{(0,0,1,0)}(\vec{v}))(\chi^{(O(8)}_{1,0,0,0)}(\vec{v}) + \chi^{(O(8)}_{(0,0,0,1)}(\vec{v})) \qquad (4.31)$$

The degenaracy of the first massive level is $\chi^{\mathrm{long}}(\vec{0}) = 16^2 = 2^8$, which is the dimension of the long multiplet of $N=1$ supersymmetry in 8 dimensions. It follows from supersymmetry that all higher massives states must fall into similar multiplets, so that $\chi_{\mathrm{GS}_-}(\vec{v},\sqrt{q_0}) - \chi_{\mathrm{GS}_-}(\vec{v},0)$ must be divisible [15] by $\chi^{\mathrm{long}}$. Let us show this explicitely. It is easy to see that, according to Eqs.4.18 and 4.20,

$$\chi^{\mathrm{long}}(\vec{v}) = \prod_\mu(\eta_\mu^{1/2}+\eta_\mu^{-1/2})\prod_\mu(\rho_\mu^{1/2}+\rho_\mu^{-1/2}),$$

so that we want to factor out these two factors. We start by writing, using Eqs.4.17, 4.19

$$\Theta_2(\vec{z},\sqrt{q_0}) - \Theta_2(\vec{y},\sqrt{q_0}) = \Theta_1(\vec{v},\sqrt{q_0})$$

Moreover, it follows from Eqs.4.18, and 4.20 that

$$\prod_\mu(\rho_\mu^{1/2}+\rho_\mu^{-1/2}) - \prod_\mu(\eta_\mu^{1/2}+\eta_\mu^{-1/2}) = \prod_\mu(q_\mu^{1/2}-q_\mu^{-1/2}).$$

These two equations show that

$$\Theta_2(\vec{y},\sqrt{q_0})\prod_\mu(q_\mu^{1/2}-q_\mu^{-1/2}) =$$

[15] This is discussed in detail for the dimensions of representations in ref.[12].

$$\Theta_2(\vec{y},\sqrt{q_0})\prod_\mu(\rho_\mu^{1/2}+\rho_\mu^{-1/2})-\prod_\mu(\eta_\mu^{1/2}+\eta_\mu^{-1/2})\,(\Theta_2(\vec{z},\sqrt{q_0})-\Theta_2(\vec{v},\sqrt{q_0}))$$

Thus we find

$$\chi_{\mathrm{GS}_-}(q_0\,|\,\vec{v})=\chi^{O(8)}_{(0,0,1,0)}(\vec{v})+\frac{\chi^{\mathrm{long}}(\vec{v})}{\Theta_1(\vec{v},\sqrt{q_0})}\left\{\frac{\Theta_2(\vec{y},\sqrt{q_0})}{\prod_\mu(\eta_\mu^{1/2}+\eta_\mu^{-1/2})}-\frac{\Theta_2(\vec{z},\sqrt{q_0})}{\prod_\mu(\rho_\mu^{1/2}+\rho_\mu^{-1/2})}\right\} \tag{4.32}$$

which is the formula we wanted to derive. A similar calculation gives

$$\chi_{\mathrm{GS}_+}(q_0\,|\,\vec{v})=\chi^{O(8)}_{(0,0,0,1)}(\vec{v})+\frac{\chi^{\mathrm{long}}(\vec{v})}{\Theta_1(\vec{v},\sqrt{q_0})}\left\{\frac{\Theta_2(\vec{y},\sqrt{q_0})}{\prod_\mu(\eta_\mu^{1/2}+\eta_\mu^{-1/2})}-\frac{\Theta_2(\vec{z},\sqrt{q_0})}{\prod_\mu(\rho_\mu^{1/2}+\rho_\mu^{-1/2})}\right\} \tag{4.33}$$

Note that the second terms should be rewritten as

$$\chi^{\mathrm{long}}(\vec{v})\frac{\prod_\mu(q_\mu^{1/2}-q_\mu^{-1/2})}{\Theta_1(\vec{v},\sqrt{q_0})}\frac{1}{\prod_\mu(q_\mu^{1/2}-q_\mu^{-1/2})}\left\{\frac{\Theta_2(\vec{y},\sqrt{q_0})}{\prod_\mu(\eta_\mu^{1/2}+\eta_\mu^{-1/2})}-\frac{\Theta_2(\vec{z},\sqrt{q_0})}{\prod_\mu(\rho_\mu^{1/2}+\rho_\mu^{-1/2})}\right\}$$

in order to pull out the overall factor in the product expansion of $\Theta_1$. It is easy to see that the factor between brackets is divisible by $\prod_\mu(q_\mu^{1/2}-q_\mu^{-1/2})$. Indeed, if any of the $v$ vanishes, the $y$'s and the $z$'s become pairwise equal up to permutation or inversion, so that the two terms inside the bracket become equal. As we will see below, this is precisely what occurs when the type II superstring theories are compactified to nine dimensions, so that $\mathrm{II}_A$ and $\mathrm{II}_B$ superstrings give the same characters since there is no chirality in nine dimensions.

# 5 Modular properties of the closed type II string characters:

## 5.1 Uncompactified target space

In terms of the variables $\vec{v}$ such that $q_\mu=\exp(2\pi i v_\mu)$, the chiral characters write

$$\chi_{\mathrm{GS}_\pm}=\frac{\prod_\mu\sin(\pi v_\mu)}{\Theta_1(\vec{v}|\tau)}\left(\frac{1}{2}\left[\Theta_3(\vec{v}|\tau)-\Theta_4(\vec{v}|\tau)\right]+\frac{1}{2}\left[\Theta_2(\vec{v}|\tau)\pm\Theta_1(\vec{v}|\tau)\right]\right).$$

Under modular transformations

$$\underline{\Theta}\left(\frac{\vec{v}}{\tau}\middle| -\frac{1}{\tau}\right) = \tau^2 e^{i\pi\vec{v}^2/\tau} \begin{pmatrix} 1 & 0 & 0 & 0 \\ 0 & 0 & 0 & 1 \\ 0 & 0 & 1 & 0 \\ 0 & 1 & 0 & 0 \end{pmatrix} \underline{\Theta}(\vec{v}|\tau) \tag{5.1}$$

Thus, as defined so far $\chi_{\mathrm{GS}_\pm}$ has no modular properties. On should define, instead,

$$\chi^{\pm}_{\mathrm{open}} = \frac{\prod_\mu \sin(\pi v_\mu)}{\Theta_1(\vec{v}|\tau)} \left(\frac{1}{2}\left[\Theta_3(\vec{v}|\tau) - \Theta_4(\vec{v}|\tau)\right] - \frac{1}{2}\left[\Theta_2(\vec{v}|\tau) \pm \Theta_1(\vec{v}|\tau)\right]\right)$$

where we put a minus sign in front of the fermionic (Ramond) contribution. Since it only contains $\Theta_2 + \Theta_4$ it has modular properties. Indeed modular invriance requires that we use the supertrace. The above discussions are only slightly modified, so we will not go through them again. Using the triality transformation displayed on Eqs.4.17, and 4.19 one sees that,

$$\chi^{+}_{\mathrm{open}} = -\frac{\prod_\mu \sin(\pi v_\mu)}{\Theta_1(\vec{v}|\tau)}\Theta_1(\vec{z}|\tau), \qquad \chi^{-}_{\mathrm{open}} = \frac{\prod_\mu \sin(\pi v_\mu)}{\Theta_1(\vec{v}|\tau)}\Theta_1(\vec{y}|\tau)$$

Note that, as a result, $\chi^{\pm}_{\mathrm{open}}$ vanishes at $\vec{v} = 0$, since then $\vec{y} = \vec{z} = 0$, and it is proportional to $\Theta_1(0|\tau)$. This is of course due to the supersymmetry.

For closed string, we multiply again by the zero mode contribution which cancels the sinus factors and get

$$\chi^{\mathrm{IIB}}_{\mathrm{closed}}(\vec{v}|\tau) \propto \frac{1}{\left|\Theta_1(\vec{v}|\tau)\right|^2}\left|\Theta_1(\vec{y}|\tau)\right|^2, \quad \text{or} \quad \frac{1}{\left|\Theta_1(\vec{v}|\tau)\right|^2}\left|\Theta_1(\vec{z}|\tau)\right|^2$$

$$\chi^{\mathrm{IIA}}_{\mathrm{closed}}(\vec{v}|\tau) \propto -\frac{\Theta_1(\vec{y}|\tau)(\Theta_1(\vec{z}|\tau))^*}{\left|\Theta_1(\vec{v}|\tau)\right|^2}, \quad \text{or} \quad \frac{\Theta_1(\vec{z}|\tau)(\Theta_1(\vec{y}|\tau))^*}{\left|\Theta_1(\vec{v}|\tau)\right|^2}.$$

It is easy to see that modular invariance holds in both cases, without multiplying by any additional factor contrary to the bosonic case. The additional factors of the theta function transformations cancel between numerators and denominators since $\vec{v}^2 = \vec{y}^2 = \vec{z}^2$.

## 5.2 Compactification on the circle

The discussion is similar to bosonic case. We are in $8 \to 7$ transverse dimensions, and $v_4 = 0$. We note $\vec{v}^{(3)} = \{v_1, v_2, v_3, 0\}$, and $\vec{y}^{(3)}$, $\vec{z}^{(3)}$ the transformed by triality. Then

$$\Theta_1(\vec{v}|\tau) \sim v_4\theta_1'(0|\tau)\prod_{\mu=1}^{3}\theta_1(v_\mu|\tau) \equiv v_4\Theta_1'(\vec{v}^{(3)}|\tau)$$

It follows from Eq.4.13 that

$$y_1 = \frac{1}{2}(v_1 - v_2 + v_3 - v_4), \quad y_2 = \frac{1}{2}(-v_1 + v_2 + v_3 - v_4),$$

$$y_3 = \frac{1}{2}(v_1 + v_2 + v_3 + v_4), \quad y_4 = \frac{1}{2}(-v_1 - v_2 + v_3 + v_4).$$

Thus, with $v_4 = 0$,

$$y_1^{(3)} = \frac{1}{2}(v_1 - v_2 + v_3), \quad y_2^{(3)} = \frac{1}{2}(-v_1 + v_2 + v_3),$$

$$y_3^{(3)} = \frac{1}{2}(v_1 + v_2 + v_3), \quad y_4^{(3)} = \frac{1}{2}(-v_1 - v_2 + v_3),$$

It follows from Eq.4.19 that

$$z_1 = \frac{1}{2}(v_1 + v_2 - v_3 + v_4), \quad z_2 = \frac{1}{2}(v_1 + v_2 + v_3 - v_4),$$

$$z_3 = \frac{1}{2}(-v_1 + v_2 + v_3 + v_4), \quad z_4 = \frac{1}{2}(v_1 - v_2 + v_3 + v_4),$$

Thus, with $v_4 = 0$,

$$z_1^{(3)} = \frac{1}{2}(v_1 + v_2 - v_3), \quad z_2^{(3)} = \frac{1}{2}(v_1 + v_2 + v_3),$$

$$z_3^{(3)} = \frac{1}{2}(-v_1 + v_2 + v_3), \quad z_4^{(3)} = \frac{1}{2}(v_1 - v_2 + v_3),$$

One sees that

$$z_1^{(3)} = -y_4^{(3)}, \quad z_2^{(3)} = y_3^{(3)}, \quad z_3^{(3)} = y_2^{(3)}, \quad z_4^{(3)} = y_1^{(3)}$$

So we find

$$\chi_{\rm open}^{+\rm circ} = \chi_{\rm open}^{-\rm circ} = \frac{\prod_{\mu=1}^{3}\sin(\pi v_\mu)}{\Theta_1'(\vec{v}^{(3)}|\tau)}\Theta_1(\vec{y}^{(3)}|\tau)$$

For closed strings, we multiply again by the zero mode factor. This gives

$$\chi_{\rm closed}^{\rm IIA,\, circ}(\tau,R|\vec{v}) = \chi_{\rm closed}^{\rm IIB,\, circ}(\tau,R|\vec{v}) = F(R,\tau)\left|\frac{(\theta_1'(0|\tau))^{2/3}}{\Theta_1'(\vec{v}^{(3)}|\tau)}\right|^2 \left|\Theta_1(\vec{x}^{(3)}|\tau)\right|^2$$

One sees that for these characters, the T duality between type IIA and type IIB superstrings in nine dimensions, corresponds to the equality between characters. This is due to the fact that chirality disappears in going from ten to nine dimensions, so that both types have the same spectrum. Concerning modularity, the product of the last two terms is modular invariant. Thus this is also true for $\chi_{\rm closed}^{\rm IIA,\, circ}(\tau,R|\vec{v})$. Moreover, clearly,

$$\chi_{\rm closed}^{\rm IIA,\, circ}(\tau,R|\vec{v}) = \chi_{\rm closed}^{\rm IIA,\, circ}(\tau,1/2R|\vec{v}).$$

The characters of types $\mathrm{II}_A$ and $\mathrm{II}_B$ are equal and thus self T-dual. It is known that this does not hold at the interacting level, where T-duality exchanges types $\mathrm{II}_A$ and $\mathrm{II}_B$.

# 6 Outlook

At this simple perturbative level, the study of duality (and triality) properties of characters has already been fruitful. The existence of a third formulation of superstrings is intriguing. Its main virtue may be to unify the two GSO projected sectors. They may be both described in terms of the same pair of world sheet fields $S^a$ and $T^a$ (or equivalently $S^{\dot{a}}$ and $T^{\dot{a}}$).

In view of the current importance of the Ramond-Ramond sector coupling, our construction of the Ramond and Neveu-Schwarz projection operators in the Green-Schwarz formalism may be useful.

The fact that the characters are invariant by T duality shows that they may be insensitive to the moduli in general, and hence a good unifying tool for string theory and its extensions.

The next problems is clearly to deal with heterotic strings and with more elaborate compactifications of the usual five types of strings. After that non perturbative states will be the next challenge. One may think that the present characters are strictly torus objects that do not allow to go beyond lowest order perturbation. However, it is possible that the final (M) theory is completely integrable in some sense, and that a sort of Bäcklund transformation to a free spectrum exists. One long reaching motivation of the present discussion is to try and derive the corresponding truly non perturbative characters by combining the present non perturbative knowledge about string theories and their extensions.

**Acknowledgements.** This work was done in part while one of us (J.-L. G) was visiting the University of Swansea with a support from the EC contract CHRXCT920069. He is grateful for the warm hospitality and enlightning discussions with Nazarov there. Discussions with D. Olive and A. Schwimmer have been very helpful.

# A Highest weight states in the bosonic string spectrum

Since this viewpoint is novel in string theories, we spell out some examples. For $D_N$, the step operators associated with the usual set of positive simple roots are

$$\begin{aligned} \mathcal{E}_\mu &= \sum_{n\geq 1} \frac{1}{n}(\zeta_{-n}^{-\mu}\zeta_n^{\mu+1} - \zeta_{-n}^{\mu+1}\zeta_n^{-\mu}), \quad 1 \leq \mu \leq N-1 \\ \mathcal{E}_N &= \sum_{n\geq 1} \frac{1}{n}(\zeta_{-n}^{-N+1}\zeta_n^{-N} - \zeta_{-n}^{-N}\zeta_n^{-N+1}). \end{aligned} \tag{A.1}$$

The associated simple roots may be written, in terms of orthonormal vectors $\vec{e}_i$ in $N$ dimensions:

$$\vec{\pi}_\mu = \vec{e}_\mu - \vec{e}_{\mu+1}, \quad \mu = 1, \cdots, N-1, \quad \vec{\pi}_N = \vec{e}_N + \vec{e}_{N+1}$$

and the fundamental representations have weights

$$\vec{\lambda}_j^{(2N)} = \sum_{k=1}^{j} \vec{e}_k, \quad j \leq N-2, \quad \vec{\lambda}_{N-1}^{(2N)} = \frac{1}{2}\left(\sum_1^{N-1} \vec{e}_k - \vec{e}_N\right), \quad \vec{\lambda}_N^{(2N)} = \sum_1^N \frac{\vec{e}_k}{2}. \tag{A.2}$$

Let us define, for later purpose

$$T_{mn} \equiv \sum_{\mu>0} :\zeta_m^\mu \zeta_n^{-\mu}: + :\zeta_n^\mu \zeta_m^{-\mu}: \tag{A.3}$$

Then it is easy to see that $T_{mn}$ commutes with all generators for arbitrary $m$, $n$. This is the invariance of the trace in the present basis. In general, a highest weight states $|\vec{\lambda}>$ is such that

$$\mathcal{H}_\mu|\vec{\lambda}>=\lambda_\mu|\vec{\lambda}>,\quad \mathcal{E}_\mu|\vec{\lambda}>=0,\quad \mu=1,\ldots,N. \tag{A.4}$$

Let us look for them level-by-level in the string spectrum. A basic point is that $\left[\zeta_r^{-1},\mathcal{E}_\mu\right]=0$, as one easily sees from Eqs.2.3. At level 1 the highest weight state is thus $\zeta_{-1}^{-1}|0>$. It satisfies

$$\mathcal{H}_\mu\zeta_{-1}^{-1}|0>=\delta_{\mu,1}\zeta_{-1}^{-1}|0>$$

Using the definition A.4, one sees that its higest weight is $\lambda_1^{(2N)}=\vec{e}_1$, which corresponds to the vector representation.

At level 2 the highest weight states are $|2,1>\equiv\zeta_{-2}^{-1}|0>$, $|2,2>\equiv\zeta_{-1}^{-1}\zeta_{-1}^{-1}|0>$, $|2,3>\equiv\sum_{\mu>0}\zeta_{-1}^{\mu}\zeta_{-1}^{-\mu}|0>$. This gives the one index, two index and zero index representations, respectively, with dimensions $2N$, $N(2N+1)-1$, $1$. The highest weights are

$$\vec{\lambda}_{|2,1>}=\vec{\lambda}_1^{(2N)},\quad \vec{\lambda}_{|2,2>}=2\vec{\lambda}_1^{(2N)},\quad \vec{\lambda}_{|2,3>}=0$$

where $\vec{\lambda}_1^{(2N)}$ is given by Eq.A.2.

At level 3 we have

$$|3,1>\equiv\zeta_{-3}^{-1}|0>,\quad |3,2>\equiv\zeta_{-2}^{-1}\zeta_{-1}^{-1}|0>,\quad |3,3>\equiv\left(\zeta_{-1}^{-1}\right)^3|0>,$$

$$|3,4>=T_{-1,-2}|0>,\quad |3,5>=T_{-1,-1}\zeta_{-1}^{-1}|0>,$$
$$|3,6>\equiv\left(\zeta_{-2}^{-2}\zeta_{-1}^{-1}-\zeta_{-2}^{-1}\zeta_{-1}^{-2}\right)|0> \tag{A.5}$$

The first five are similar to the previous one. The last one is of a novel nature as it involves $\zeta_{-r}^{-2}$. The only non trivial part in checking Eq.A.4 is for $\mathcal{E}_1$, since $[\zeta_{-r}^{-2},\mathcal{E}_\mu]=-\delta_{\mu,1}\zeta_{-r}^{-1}$. The total result is zero by antisymmetry. The highest weights are given by

$$\vec{\lambda}_{|3,1>}=\vec{\lambda}_1^{(2N)},\quad \vec{\lambda}_{|3,2>}=2\vec{\lambda}_1^{(2N)},\quad \vec{\lambda}_{|3,3>}=3\vec{\lambda}_1^{(2N)}$$

$$\vec{\lambda}_{|3,4>}=0,\quad \vec{\lambda}_{|3,5>}=\vec{\lambda}_1^{(2N)},$$
$$\vec{\lambda}_{|3,6>}=\vec{\lambda}_2^{(2N)}, \tag{A.6}$$

where $\vec{\lambda}_i$ is given by Eq.A.2.

Clearly, the higher levels are discussed in a similar way. They are constructed from

$$\zeta_{-p}^{-1}, \quad T_{-p_1,-p_2}, \quad F^{(n)}_{p_1,\cdots p_n} = \sum_\sigma (-1)^\sigma \zeta^{-n}_{-\sigma(p_n)} \cdots \zeta^{-1}_{-\sigma(p_1)}, \tag{A.7}$$

applied to the vacuum. The latter gives the highest weight $\sum_r p_r$. Thus the highest weights $\lambda_p^{(2N)}$ appears for the level $L \geq \sum_1^p p = p(p+1)/2$. Therefore, in Eq.A.4, the commutator with $\mathcal{E}_N$ is not trivially zero only at level $L \geq (N-1)N/2$. For the critical case $N=12$, this gives $L \geq 66$!

In general one sees that the search for highest weight states is done by arranging the transverse modes is a non covariant way which is very unusual in string theory.

# B Derivation of the Jacobi identities

Using the series expansion of $\theta$ functions one finds, for arbitrary argument noted $\vec{x}$,

$$\Theta_2(\vec{x}, \sqrt{q_0}) \pm \Theta_1(\vec{x}, \sqrt{q_0}) = \sum_{m_\mu \text{ half integers}} (1 \pm e^{i\pi \sum_\mu m_\mu}) q_0^{\sum_\mu (m_\mu)^2/2} e^{i\pi \sum_\mu 2(m_\mu) x_\mu}$$

$$\Theta_3(\vec{x}, \sqrt{q_0}) \pm \Theta_4(\vec{x}, \sqrt{q_0}) = \sum_{m_\mu \text{ integers}} (1 \pm e^{i\pi \sum_\mu m_\mu}) q_0^{\sum_\mu (m_\mu)^2/2} e^{i\pi \sum_\mu 2(m_\mu) x_\mu}$$

Clearly, $\sum_\mu m_\mu$ is integer. So the factors $(1 \pm e^{i\pi \sum_\mu m_\mu})$ forces it to be even or odd, respectively, and we may write

$$\frac{1}{2}\left(\Theta_2(\vec{x}, \sqrt{q_0}) - \Theta_1(\vec{x}, \sqrt{q_0})\right) = \sum_{\substack{m_\mu \text{ half integers,} \\ \sum_\mu m_\mu \text{ odd}}} q_0^{\sum_\mu (m_\mu)^2/2} e^{i\pi \sum_\mu 2(m_\mu) x_\mu}$$

$$\frac{1}{2}\left(\Theta_2(\vec{x}, \sqrt{q_0}) + \Theta_1(\vec{x}, \sqrt{q_0})\right) = \sum_{\substack{m_\mu \text{ half integers,} \\ \sum_\mu m_\mu \text{ even}}} q_0^{\sum_\mu (m_\mu)^2/2} e^{i\pi \sum_\mu 2(m_\mu) x_\mu}$$

$$\frac{1}{2}\left(\Theta_3(\vec{x}, \sqrt{q_0}) - \Theta_4(\vec{x}, \sqrt{q_0})\right) = \sum_{\substack{m_\mu \text{ integers,} \\ \sum_\mu m_\mu \text{ odd}}} q_0^{\sum_\mu (m_\mu)^2/2} e^{i\pi \sum_\mu 2(m_\mu) x_\mu}$$

$$\frac{1}{2}\left(\Theta_3(\vec{x}, \sqrt{q_0}) + \Theta_4(\vec{x}, \sqrt{q_0})\right) = \sum_{\substack{m_\mu \text{ integers,} \\ \sum_\mu m_\mu \text{ odd}}} q_0^{\sum_\mu (m_\mu)^2/2} e^{i\pi \sum_\mu 2(m_\mu) x_\mu} \tag{B.1}$$

## The q (or v) to eta (or y) transformation

Let $\vec{x} = \vec{v}$ in the above formula. In the summations, we make the change of variables $\vec{n} = \mathcal{M}.\vec{m}$. Then,

$$\sum_\mu m_\mu x_\mu = \sum_\mu n_\mu y_\mu \quad \sum_\mu m_\mu^2 = \sum_\mu n_\mu^2.$$

Consider the range of summations. Explicitly, one has

$$n_1 = \frac{1}{2}(m_1 - m_2 + m_3 - m_4), \quad n_2 = \frac{1}{2}(-m_1 + m_2 + m_3 - m_4),$$

$$n_3 = \frac{1}{2}(m_1 + m_2 + m_3 + m_4), \quad n_4 = \frac{1}{2}(-m_1 - m_2 + m_3 + m_4).$$

By construction, the $m_i \pm m_j$'s are integers. One easily sees that $n_i \pm n_j$ are integers. Thus the $n$'s are all integers or half integers simultaneously. Looking at the above formulae, it is chear that $n_3$ is integer if $\sum_\mu m_\mu$ is even, and half integer if $\sum_\mu m_\mu$ is odd. Thus the same is true for all $n$'s. Moreover, since $\mathcal{M}^2 = 1$, $\sum_\mu n_\mu = 2m_3$. Thus it is even for integer $m$'s, and odd for half integer $m$'s. The correspondence may be summarised as follows:

| Theta(v) | $\sum_\mu m_\mu$ | rge of m | $\sum_\mu n_\mu$ | rge of n | Theta(y) |
|---|---|---|---|---|---|
| $\Theta_3 + \Theta_4$ | even | integer | even | integer | $\Theta_3 + \Theta_4$ |
| $\Theta_3 - \Theta_4$ | odd | integer | even | half integer | $\Theta_2 + \Theta_1$ |
| $\Theta_2 + \Theta_1$ | even | half integer | odd | integer | $\Theta_3 - \Theta_4$ |
| $\Theta_2 - \Theta_1$ | odd | half integer | odd | half integer | $\Theta_2 - \Theta_1$ |

This is equivalent to Eqs.4.17.

## The q (or v) to rho (or z) transformation

Similarly, one lets $\vec{p} = \mathcal{N}\vec{m}$, so that

$$\sum_\mu m_\mu x_\mu = \sum_\mu p_\mu z_\mu \quad \sum_\mu m_\mu^2 = \sum_\mu p_\mu^2$$

One has, according to Eq.4.19

$$p_1 = \frac{1}{2}(m_1 + m_2 - m_3 + m_4), \quad p_2 = \frac{1}{2}(m_1 + m_2 + m_3 - m_4),$$

$$p_3 = \frac{1}{2}(-m_1 + m_2 + m_3 + m_4), \quad p_4 = \frac{1}{2}(m_1 - m_2 + m_3 + m_4).$$

which may be rewritten as

$$p_1 = -m_3 + \frac{1}{2}(\sum_\mu m_\mu), \quad p_2 = -m_4 + \frac{1}{2}(\sum_\mu m_\mu),$$

$$p_3 = -m_1 + \frac{1}{2}(\sum_\mu m_\mu), \quad p_4 = -m_2 + \frac{1}{2}(\sum_\mu m_\mu).$$

Clearly the $p$'s are all integer or half-integer simultaneously. If $\sum_\mu m_\mu$ is even (resp. odd) they are integer or half integer at the same time (resp. at opposite time) as the $m$'s. On the other hand $\sum_\mu p_\mu = \sum_\mu m_\mu$. The correspondence is summarised as follows:

| Theta(v) | $\sum_\mu m_\mu$ | rge of m | $\sum_\mu p_\mu$ | rge of p | Theta(z) |
|---|---|---|---|---|---|
| $\Theta_3 + \Theta_4$ | even | integer | even | integer | $\Theta_3 + \Theta_4$ |
| $\Theta_3 - \Theta_4$ | odd | integer | odd | half integer | $\Theta_2 - \Theta_1$ |
| $\Theta_2 + \Theta_1$ | even | half integer | even | half integer | $\Theta_2 + \Theta_1$ |
| $\Theta_2 - \Theta_1$ | odd | half integer | odd | integer | $\Theta_3 - \Theta_4$ |

This is equivalent to Eq.4.19.

# C The fermion-boson aspect

In this appendix[16] we show how our extended framework appears in the usual fermionisation scheme. The four summations of Eqs.B.1 are of course, the well known four equivalence classes of the weight lattice of $O(8)$. We denote them by $\Lambda^{(4)o,h}$, $\Lambda^{(4)e,h}$, $\Lambda^{(4)o,i}$, $\Lambda^{(4)e,i}$, where $e$ $o$ refer to whether $\sum_\mu m_\mu$ is even or odd, and $i$ and $h$ distinguish the case of integer and half integer components. In view of the summary of section 4.5, and looking at Eqs.B.1 one sees that we have the correspondence:

| Theta(v) | $\Theta_3 + \Theta_4$ | $\Theta_3 - \Theta_4$ | $\Theta_2 + \Theta_1$ | $\Theta_2 - \Theta_1$ |
|---|---|---|---|---|
| sector | $NS_-$ | $NS_+$ | $R_+$ | $R_-$ |
| equiv. class | $\Lambda^{(4)e,i}$ | $\Lambda^{(4)o,i}$ | $\Lambda^{(4)e,h}$ | $\Lambda^{(4)o,h}$ |

[16]This section was written after a stimulating remark of A. Schwimmer

Define on this zero mode lattice a bosonic vertex operator

$$V_{\vec{u}}(z) = e^{i\vec{u}\vec{q}_0} z^{\vec{\alpha}_0 \vec{u}} z^{\vec{u}^2/2} \exp\left(-\sum_{n<0} \frac{\vec{u}\vec{\alpha}_n}{n} z^{-n}\right) \exp\left(-\sum_{n>0} \frac{\vec{u}\vec{\alpha}_n}{n} z^{-n}\right)$$

If $\vec{u}^2 = 1$ this operator is a world sheet fermion. There are three ways to obtain such vectors of length one.

**Choose $\vec{u}$ with integral components** This means that $\vec{u} = \vec{e}_i$ for some $i$, where $\vec{e}_i$ is one of the basis vectors. Then, as is well known[3], $V_{\vec{u}}(z)$ has integer modes in $\Lambda^{(4)h}$ (Ramond sector where $V_{\pm\vec{e}_\mu}(z)) \equiv \Gamma_{\pm\mu}(z)$) and half integer modes in $\Lambda^{(4)i}$ (NS sector where $V_{\pm\vec{e}_\mu}(z)) \equiv \psi_\mu(z)$). Clearly

$$V_{\vec{e}_\mu}(z)\Lambda^{(4)e,i} \in \Lambda^{(4)o,i}, \quad V_{\vec{e}_\mu}(z)\Lambda^{(4)o,i} \in \Lambda^{(4)e,i}$$

$$V_{\vec{e}_\mu}(z)\Lambda^{(4)e,h} \in \Lambda^{(4)o,h}, \quad V_{\vec{e}_\mu}(z)\Lambda^{(4)o,h} \in \Lambda^{(4)e,h}$$

**Choose $\vec{u}$ with half integral components** Now, $V_{\pm\vec{u}_\mu}(z)$ shifts $\vec{\alpha}_0$ by half integers, so that it exchanges $\Lambda^{(4)i}$ and $\Lambda^{(4)h}$ There are two possible choices:

**even number of minus signs:** using the unitary symmetric matrix $\mathcal{M}$ we derive the orthonormal vectors

$$\vec{u}_\mu^{\mathcal{M}} = \sum_\nu \mathcal{M}_{\mu\nu}\vec{e}_\nu$$

Now

$$V_{\vec{u}_\mu^{\mathcal{M}}}(z)\Lambda^{(4)e,i} \in \Lambda^{(4)e,h}, \quad V_{\vec{u}_\mu^{\mathcal{M}}}(z)\Lambda^{(4)o,i} \in \Lambda^{(4)o,h}$$

$$V_{\vec{u}_\mu^{\mathcal{M}}}(z)\Lambda^{(4)e,h} \in \Lambda^{(4)e,i}, \quad V_{\vec{u}_\mu^{\mathcal{M}}}(z)\Lambda^{(4)o,h} \in \Lambda^{(4)o,i}$$

that is

$$V_{\vec{u}_\mu^{\mathcal{M}}}(z)NS_- \in R_+, \quad V_{\vec{u}_\mu^{\mathcal{M}}}(z)NS_+ \in R_-$$

$$V_{\vec{u}_\mu^{\mathcal{M}}}(z)R_+ \in NS_-, \quad V_{\vec{u}_\mu^{\mathcal{M}}}(z)R_- \in NS_+$$

Thus one sees that on $\mathrm{NS}_+$ and $\mathrm{R}_-$, we have $V_{\vec{u}^{\mathcal{M}}} \equiv S^a$, which is the well known way[3] to understand the relationship between NSR and GS formulation from the fermion-boson correspondence. On the other hand, on the two other sectors $\mathrm{NS}_-$ and $\mathrm{R}_+$ $V_{\vec{u}^{\mathcal{M}}} \equiv T^a$.

**odd number of minus signs:** using the unitary symmetric matrix $\mathcal{N}$ we derive the orthonormal vectors

$$\vec{u}^{\mathcal{N}}_{\mu} = \sum_{\nu} \mathcal{N}_{\mu\nu} \vec{e}_{\nu}$$

$$V_{\vec{u}^{\mathcal{N}}_{\mu}}(z)\Lambda^{(4)e,i} \in \Lambda^{(4)o,h}, \quad V_{\vec{u}^{\mathcal{N}}_{\mu}}(z)\Lambda^{(4)o,i} \in \Lambda^{(4)e,h}$$

$$V_{\vec{u}^{\mathcal{N}}_{\mu}}(z)\Lambda^{(4)e,h} \in \Lambda^{(4)o,i}, \quad V_{\vec{u}^{\mathcal{N}}_{\mu}}(z)\Lambda^{(4)o,h} \in \Lambda^{(4)e,i}$$

that is

$$V_{\vec{u}^{\mathcal{N}}_{\mu}}(z)NS_{-} \in R_{-}, \quad V_{\vec{u}^{\mathcal{N}}_{\mu}}(z)NS_{+} \in R_{+}$$

$$V_{\vec{u}^{\mathcal{N}}_{\mu}}(z)R_{+} \in NS_{+}, \quad V_{\vec{u}^{\mathcal{N}}_{\mu}}(z)R_{-} \in NS_{-}$$

Thus one sees that on NS$_+$ and R$_+$, we have $V_{\vec{u}^{\mathcal{N}}} \equiv S^{\dot{a}}$, which is the well known way[3] to understand the relationship between NSR and GS formulation from the fermion-boson correspondence. On the other hand, in the two other sectors, NS$_-$ and R$_-$ $V_{\vec{u}^{\mathcal{N}}} \equiv T^{\dot{a}}$.

The features summarised in the present appendix are represented in the following figures where the two GSO projections are displayed

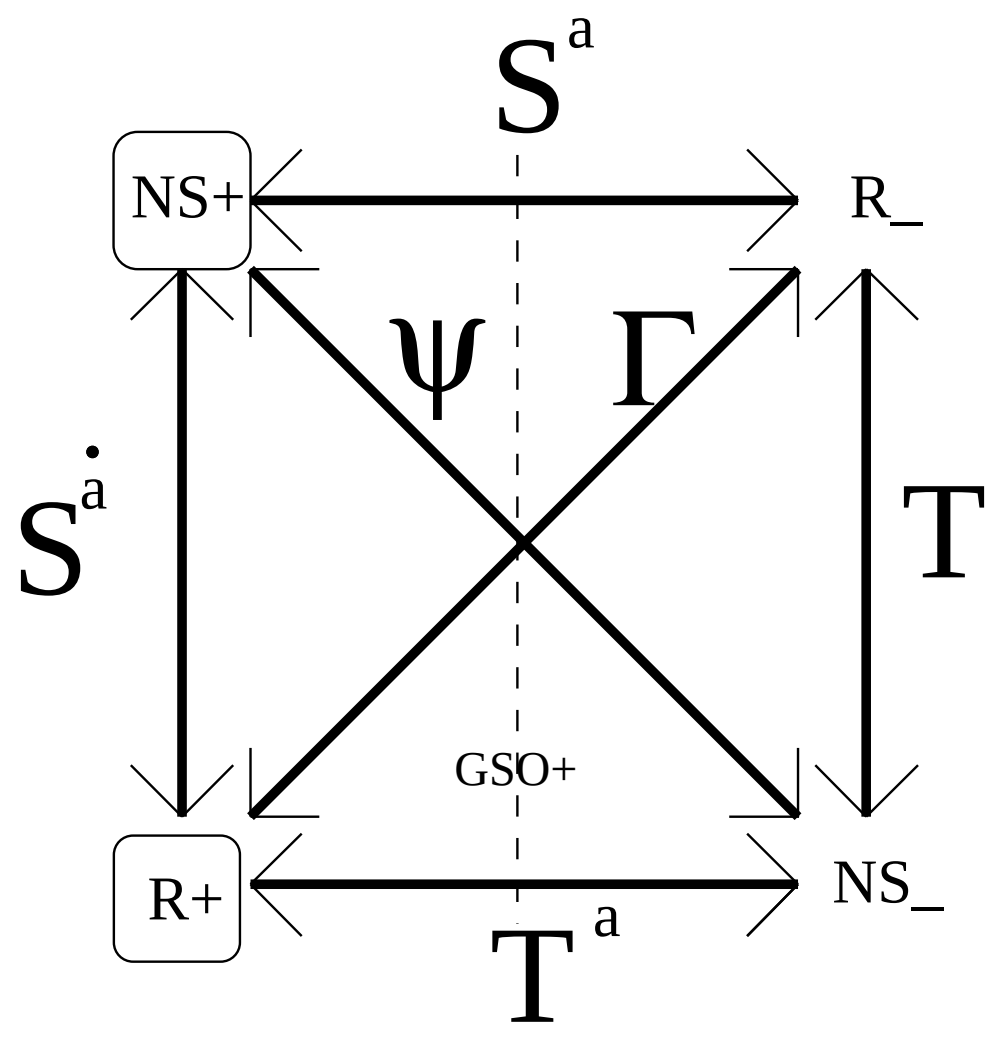

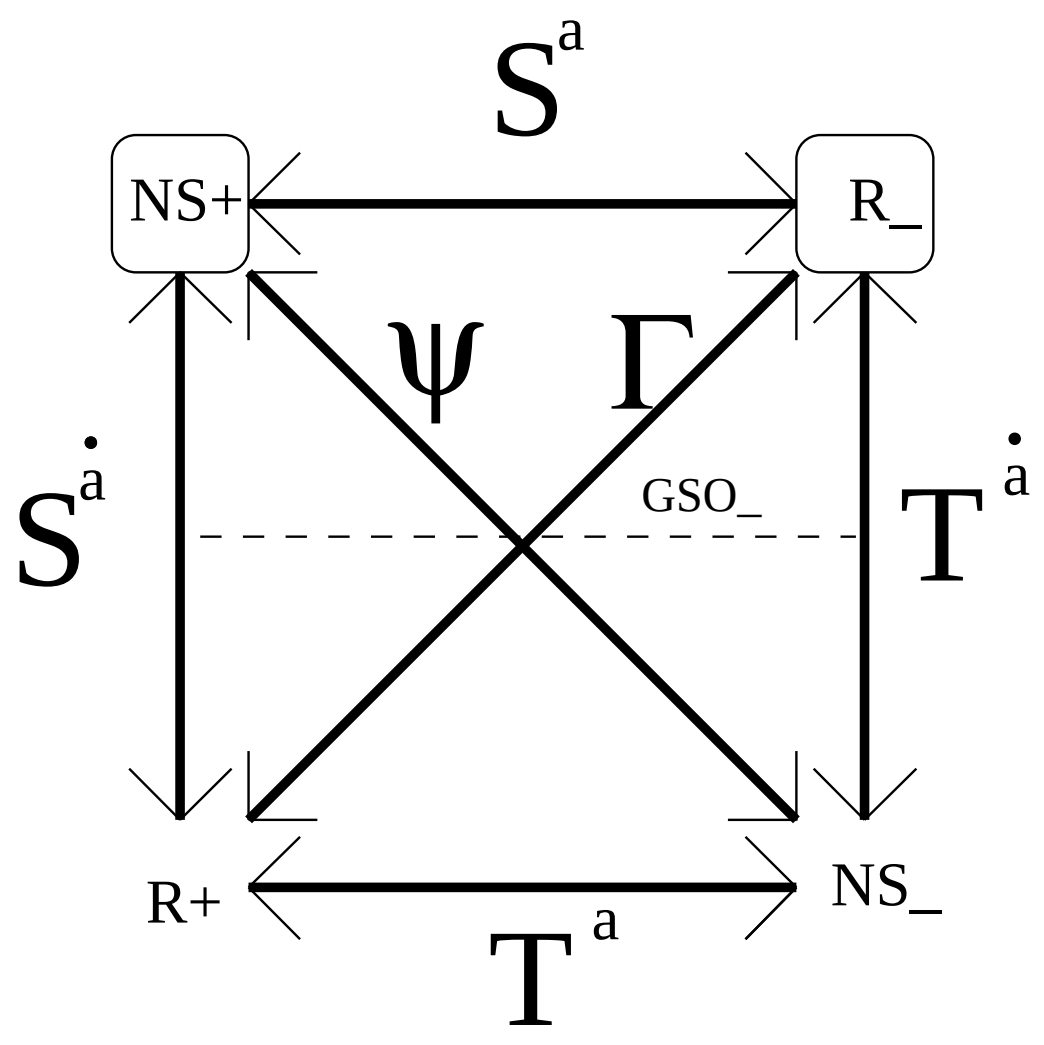